\documentclass[lettersize,12pt,draftclsnofoot,onecolumn]{IEEEtran}
\usepackage{amsmath,amsfonts}
\usepackage{cite}
\usepackage{algorithmic}
\usepackage{algorithm}
\usepackage{array}
\usepackage[caption=false,font=normalsize,labelfont=sf,textfont=sf]{subfig}
\usepackage{textcomp}
\usepackage{stfloats}
\usepackage{url}
\usepackage{verbatim}
\usepackage{graphicx}
\usepackage{cite}
\usepackage{multirow}
\usepackage{xcolor}

\usepackage{booktabs}

\usepackage{booktabs}
\usepackage{makecell}

\begin{document}

\bstctlcite{BSTcontrol}

\title{Multi-Step Underwater Acoustic Channel Prediction via Dual-Domain Structure Preservation and Pretrained Transformer}

\author{Mingjie Cao$^{\dagger}$,~\IEEEmembership{Member,~IEEE,}
        Gewei~Zhang$^{\dagger}$,~\IEEEmembership{Member,~IEEE,}
        Deqing~Wang,~\IEEEmembership{Member,~IEEE,}
        Lizhao~You,~\IEEEmembership{Member,~IEEE,}
        Fei-Yun~Wu,~\IEEEmembership{Senior~Member,~IEEE,}
        and Liqun~Fu,~\IEEEmembership{Senior~Member,~IEEE}
\thanks{$^{\dagger}$These authors contributed equally to this work.}
\thanks{M. Cao, G. Zhang, D. Wang, L. You, and L. Fu are with the
Department of Information and Communication Engineering,
School of Informatics, Xiamen University, Xiamen 361005, China
(e-mail: caomingjie@stu.xmu.edu.cn;
zhanggewei@stu.xmu.edu.cn;
deqing@xmu.edu.cn;
lizhaoyou@xmu.edu.cn;
liqun@xmu.edu.cn).}

\thanks{D. Wang is also with the Key Laboratory of Underwater Acoustic
Communication and Marine Information Technology (Xiamen University),
Ministry of Education, Xiamen 361005, China.}

\thanks{F.-Y. Wu is with the Navigation College, Jimei University,
Xiamen, Fujian 361021, China
(e-mail: feiyun@jmu.edu.cn).}

\thanks{Corresponding author: Deqing Wang.}
}



\maketitle

\pagestyle{plain}
\thispagestyle{plain}
\begin{abstract}
Accurate future channel state information (CSI) is essential for underwater acoustic (UWA) communication because delays associated with signal propagation, feedback, scheduling, and transceiver reconfiguration can render currently available CSI outdated. However, most existing predictors estimate only the next channel state and obtain multi-step forecasts through recursive rollout, which causes prediction errors to accumulate as the multipath structure evolves. The scarcity of sea-trial measurements further limits reliable model training. To address these challenges, we propose LLM4UACP, a dual-domain, structure-preserving framework that adapts a pretrained Transformer for few-shot multi-step UWA channel prediction. LLM4UACP extracts complementary features from the channel impulse response (CIR) and channel frequency response (CFR), aligns them through Fourier-consistent fusion, and encodes the fused representation as delay-time tokens. A pretrained Transformer with a frozen backbone is adapted through Butterfly Orthogonal Fine-tuning to jointly predict multiple future CIRs in a single forward pass, thereby avoiding the error accumulation associated with recursive prediction. Experiments on the measured SO-06 and SO-07 channels show that LLM4UACP achieves normalized mean squared error (NMSE) values of 0.0411 and 0.0399, corresponding to reductions of 88.53\% and 85.15\%, respectively, relative to the most recent baseline. With only 0.11M trainable parameters, LLM4UACP maintains robust performance across varying prediction horizons and observation-window lengths, limited-data settings, and cross-dataset transfer scenarios.
\end{abstract}


\section{Introduction}
\label{sec:intro}

\IEEEPARstart{U}{nderwater} acoustic (UWA) communication  and networking support a wide range of marine applications, including ocean monitoring, offshore exploration, underwater vehicle coordination, and distributed sensing networks \cite{jahanbakht2021internet,Hou2023EnvironmentAware,yang2024Review,li2025nature}. Timely and accurate acquisition of channel state information (CSI), particularly the channel impulse response (CIR), is crucial for coherent detection, adaptive equalization, and feedback-driven transmission adaptation \cite{zhang2020efficient,24,23,18,liang2024joint}. However, UWA channels are inherently doubly selective in time and frequency due to multipath propagation, Doppler shifts and spreading caused by relative platform motion and surface dynamics, long delay spreads, and ocean-medium inhomogeneity \cite{qarabaqi2013statistical,zhao2016parameter,yan2026highspeed,men2025ofdm}. Consequently, the CIR may vary substantially even over a short time interval, and a CIR estimated at a previous instant may not accurately characterize the channel at a subsequent instant. Such channel aging can impair coherent reception and adaptive equalization, and further degrade link reliability and network-level adaptation. Therefore, accurate prediction of future UWA CIRs is essential for robust underwater communication and network adaptation. Nevertheless, this problem is highly challenging, since the CIR is governed by rapidly evolving multipath structures, Doppler-induced time scaling, and dynamic ocean conditions, which jointly produce complex and nonstationary channel evolution that is difficult to extrapolate over multiple future steps~\cite{liu2024multi}.

\subsection{Related work}

Existing UWA channel prediction methods can be broadly categorized into linear recursive filtering, model-based tracking, and data-driven learning. Linear recursive predictors update prediction coefficients from recent channel observations, enabling online operation with modest computational cost \cite{radosevic2014adaptiveofdm,20,zhang2021adaptive}. However, they treat the sampled CIR as independent tap-wise sequences and therefore reduce channel prediction to local coefficient extrapolation. Path-level dynamics, including delay migration, Doppler-induced phase rotation, and path birth or death, instead appear as irregular variations in tap amplitude, phase, and support of discrete delay taps. Their performance is consequently sensitive to the predictor order and observation window, particularly when the channel evolves beyond the captured local pattern.

Model-based methods address the limitations of coefficient-wise tracking by imposing explicit temporal structure on the channel state. Kalman-type filters and forward--backward predictors commonly use autoregressive models for recursive UWA channel tracking \cite{21,31,huang2021dynamic}. Sparse methods track only the dominant delay taps \cite{li2007estimation}, whereas subspace methods track correlated components in a low-dimensional space \cite{10,huang2013model,haghighatshoar2018low}. These structural assumptions can improve tracking performance, but the methods depend on prespecified state-transition models, delay supports, or subspaces. In time-varying underwater acoustic channels, violations of these assumptions can bias the channel estimates and cause errors to accumulate across recursive updates.

Data-driven methods relax the fixed structural assumptions of model-based approaches by learning channel dynamics directly from observations. CsiPreNet combines a 1-D CNN with an LSTM to predict frequency-domain CSI for adaptive UWA OFDMA systems \cite{csiprenet}. However, because it operates only in the channel frequency response (CFR)\cite{huang2016channel} domain, it does not explicitly account for the locality and sparsity of multipath arrivals in the delay domain. AttLSTMPreNet introduces temporal attention \cite{attlstm}, while ABiGRU uses an adaptive bidirectional GRU for online CIR prediction \cite{6}. Although these recurrent models improve temporal modeling, extending them to multi-step prediction typically requires recursive rollout, which may lead to error accumulation as channel dynamics evolve.

Recent neural predictors go beyond generic sequence fitting by introducing structured learning targets for UWA channel prediction. The multi-task framework in \cite{multi-task} decomposes high-dimensional CIR prediction into correlated delay-segment tasks and uses shared feature learning (SFL) to model cross-path dependencies, reducing full-CIR prediction complexity. However, path energy crossing predefined segment boundaries may disrupt continuous path evolution and phase-coherent CIR reconstruction. Delay--Doppler ConvLSTM predicts the next channel matrix from sparse, relatively stable delay--Doppler representations \cite{jing2026delay}, using dominant Doppler indices selected from the first historical matrix and fixed during prediction. This assumes stable dominant components within the prediction window. Both methods improve prediction through structured representations but depend on their consistency with path evolution.

\subsection{Motivation and Contributions}

The above methods improve UWA channel prediction from different perspectives, including coefficient tracking, model-based state evolution, and data-driven temporal learning. \textbf{Nevertheless, two practical challenges remain insufficiently addressed for UWA links: reliable multi-step channel prediction when the currently estimated CSI becomes outdated before use, and robust learning from scarce and deployment-specific sea-trial measurements.}
First, many existing predictors are designed primarily for one-step prediction. In practical UWA links, a non-negligible interval may exist between channel prediction and its actual use due to propagation, feedback, scheduling, and signal-processing delays. During this interval, the channel may evolve substantially, causing the previously predicted CSI to deviate from the channel state at the time of use. Predicting the channel over the resulting acquisition-to-use interval therefore requires channel states several steps ahead. Applying one-step predictors over such horizons commonly relies on recursive rollout, where each predicted CIR is reused to generate the next prediction. The resulting errors can propagate across prediction steps as path gains, propagation delays, and active multipath components evolve~\cite{Jiang2022Accurate}. Second, reliable sea-trial measurements are costly to collect, limited in scale, and strongly deployment-specific. Training large neural predictors from scratch with such data can lead to overfitting and poor generalization~\cite{zhuang2024diffusion}. These challenges motivate a data-efficient framework that jointly predicts multiple future CIRs in a single forward pass, thereby avoiding recursive error accumulation.

To directly address these issues, we propose LLM4UACP, a parameter-efficient framework for few-shot multi-step UWA channel prediction. \textbf{LLM4UACP predicts multiple future CIRs in a single forward pass, avoiding the error accumulation inherent in recursive one-step rollout.} To learn effectively from scarce sea-trial data, it reuses a pretrained time-series Transformer as the sequence-modeling backbone and adapts it with Butterfly Orthogonal Fine-tuning (BOFT) while keeping most parameters frozen~\cite{boft}. Here, the Transformer is used for its long-context self-attention and reusable sequence-modeling capacity. Unlike structured neural predictors such as the multi-task learning framework in~\cite{multi-task}, which imposes predefined segment-level tasks on the CIR, LLM4UACP maps dual-domain channel observations into structure-preserving delay--time tokens and models their multi-step evolution in a unified token space. This design reduces dependence on fixed structural partitions and better captures path-level channel changes. A linear bypass further preserves local channel dynamics, while a temporal-gradient consistency loss regularizes CIR evolution across prediction steps.

The main contributions of this paper are summarized as follows:
\begin{itemize}
    \item We propose a dual-domain, structure-preserving representation for multi-step UWA CIR prediction, where CIR and CFR features are aligned in a delay-indexed latent space and tokenized into delay--time patches for Transformer-based sequence modeling.

    \item We introduce a parameter-efficient adaptation strategy based on BOFT, which updates only structured orthogonal butterfly transformations while keeping the pretrained Transformer frozen, enabling effective learning from limited sea-trial data.

    \item We evaluate LLM4UACP on measured sea-trial UWA channels under multi-step prediction settings, demonstrating its prediction accuracy, horizon robustness, and parameter efficiency against representative statistical and learning-based baselines.

\end{itemize}

The remainder of this paper is organized as follows. Section~\ref{sec:system_model} introduces the UWA channel model and multi-step prediction task. Section~\ref{sec:dataset} describes the sea-trial datasets and channel characteristics. Section~\ref{sec:framework} presents the proposed LLM4UACP framework. Section~\ref{sec:training} gives the training setup and evaluation metrics. Section~\ref{sec:expreiment} reports the experimental results and ablation studies. Section~\ref{sec:conclusion} concludes the paper.

\section{System Model and Problem Formulation}
\label{sec:system_model}

UWA channels are mainly characterized by multipath propagation and Doppler-induced time variation. To describe this structure at the $n$-th channel observation instant, the time-varying UWA channel is characterized by its CIR $h_n(\tau)$. Accounting for the distinct multipath propagation and environmental dynamics, the multipath structure of $h_n(\tau)$ is expressed as:
\begin{equation}
h_n(\tau)=\sum_{\ell=1}^{L_n}\alpha_{\ell,n}\psi(\tau-\tau_{\ell,n})+\eta_n(\tau),
\label{eq:continuous_multipath}
\end{equation}
where $L_n$ is the number of resolvable paths, $\alpha_{\ell,n}$ and $\tau_{\ell,n}$ denote the complex gain and delay of the $\ell$-th path, respectively, $\psi(\cdot)$ represents the effective pulse shape, and $\eta_n(\tau)$ models the residual unmodeled components. The UWA CIR evolves as path gains, propagation delays, and active multipath components change over time. These physical variations appear in the observed CIR as amplitude fluctuations, phase variations, delay migration, and support changes.

\begin{figure}[!t]
\centering
\includegraphics[width=0.8\columnwidth]{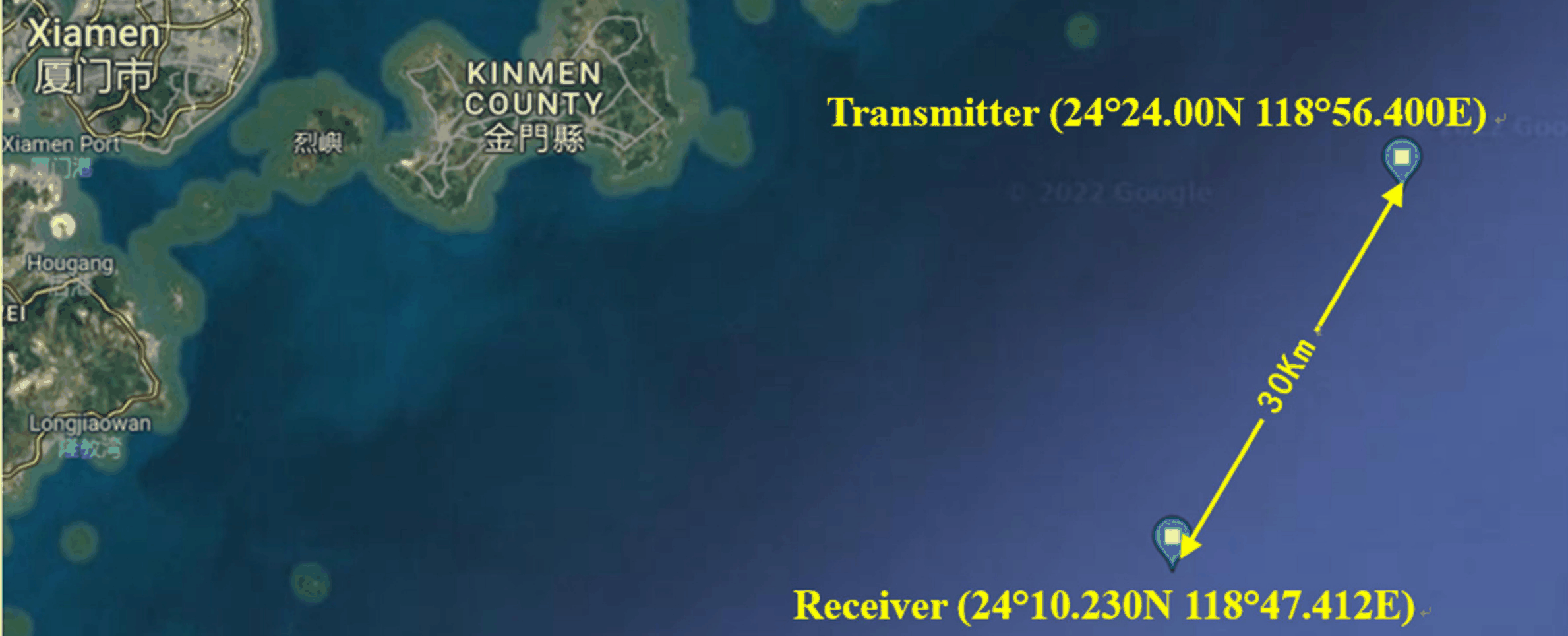}
\caption{Sea-trial measurement geometry in the Taiwan Strait.}
\label{fig:sea_trial}
\end{figure}

\begin{figure}[!t]
\centering
\includegraphics[width=0.9\columnwidth]{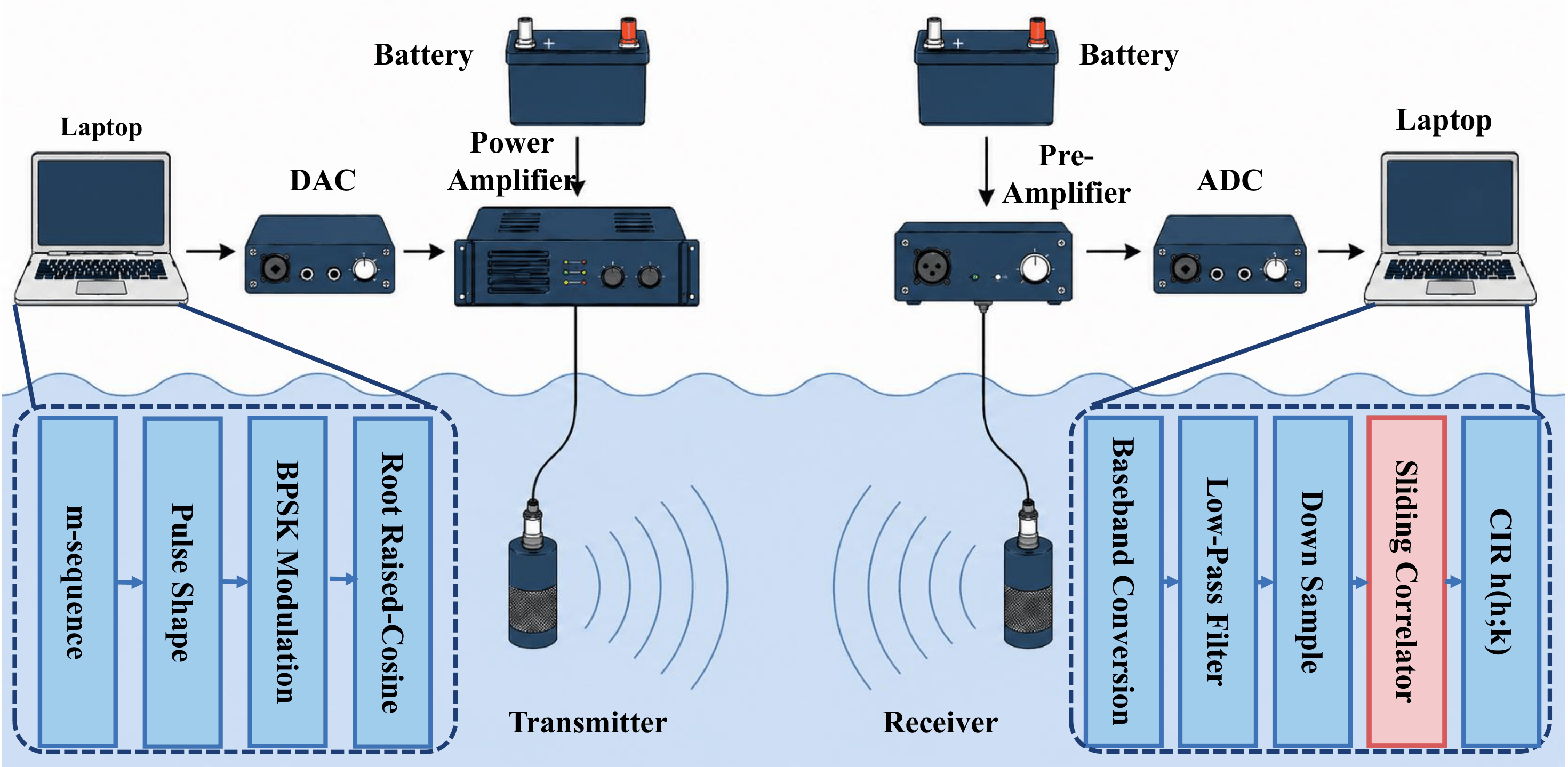}
\caption{Experimental setup of the underwater acoustic channel sounding sea trial, including the DAC-based transmission chain, ADC-based reception chain, and baseband signal processing procedure.}
\label{fig:channelSounding}
\end{figure}

Sampling the delay axis with a uniform spacing $\Delta_{\tau}$ yields the discrete CIR vector $\mathbf{h}_{n}=[h_{n,1},h_{n,2},\ldots,h_{n,K}]^{T}\in\mathbb{C}^{K}$, where $K$ denotes the number of delay taps. Over a historical observation window of length $T_{\mathrm{obs}}$ , the accumulated delay-domain observation matrix is constructed as:
\begin{equation}
\mathbf{H}_{\mathrm{obs}}^{D}
= \bigl[\mathbf{h}_{n-T_{\mathrm{obs}}+1},\dots,\mathbf{h}_{n}\bigr]^{\!T}
\in\mathbb{C}^{T_{\mathrm{obs}}\times K}.
\label{eq:delay_observation_block}
\end{equation}
Here, $T_{\mathrm{obs}}$ denotes the number of consecutive CIR snapshots rather than the duration of a communication symbol.

To obtain a frequency-domain view of the same channel, a DFT is
applied along the delay dimension. Let $\mathbf{F}_{K}$ be the
$K$-point orthonormal DFT matrix,
\begin{equation}
[\mathbf{F}_{K}]_{m,k} = \frac{1}{\sqrt{K}} \exp\left(-j\frac{2\pi}{K}(m-1)(k-1)\right),
\label{eq:dft_matrix}
\end{equation}
where $m,k = 1,2,\dots,K$. The transformed vector is:
\begin{equation}
\mathbf{g}_n=\mathbf{F}_{K}\mathbf{h}_n \in \mathbb{C}^{K}.
\end{equation}
Here, $\mathbf{g}_n$ is the CFR
vector at the $n$-th observation instant. Each entry corresponds
to one discrete frequency bin. The CFR vectors in the same observation window are stacked as:
\begin{equation}
\mathbf{H}^{F}_{\rm obs}=[\mathbf{g}_{n-T_{\rm obs}+1},\mathbf{g}_{n-T_{\rm obs}+2},\ldots,\mathbf{g}_{n}]^T \in \mathbb{C}^{T_{\rm obs}\times K}.
\end{equation}

These dual-domain representations provide aligned yet distinct physical insights. Specifically, $\mathbf{H}_{\mathrm{obs}}^{D}$ captures sparse multipath arrivals and temporal variations, whereas $\mathbf{H}_{\mathrm{obs}}^{F}$ delineates the frequency-selective fading induced by multipath interference. Although the final prediction target is the future CIR trajectory, the CFR complements the delay-domain representation by revealing multipath-induced frequency selectivity and spectral coherence.

We formulate UWA channel prediction as a dual-domain-to-delay-domain
multi-step forecasting problem. In such links, long propagation delays and processing latency can render the current CIR obsolete before it is used, making multi-step prediction essential for timely transmission adaptation. Given the historical dual-domain observations
over $T_{\mathrm{obs}}$ snapshots,
\begin{equation}
\mathcal{X}_{n}
=
\left\{
(\mathbf{h}_{n-T_{\mathrm{obs}}+1},\mathbf{g}_{n-T_{\mathrm{obs}}+1}),
\dots,
(\mathbf{h}_{n},\mathbf{g}_{n})
\right\},
\end{equation}
the goal is to predict the future delay-domain CIR sequence over the
next $T_{\mathrm{pred}}$ snapshots:
\begin{equation}
\mathbf{H}_{\mathrm{target}}^{D}
=
\left[
\mathbf{h}_{n+1},
\mathbf{h}_{n+2},
\dots,
\mathbf{h}_{n+T_{\mathrm{pred}}}
\right]^{T}
\in \mathbb{C}^{T_{\mathrm{pred}}\times K}.
\label{eq:prediction_tensor}
\end{equation}
Here, $T_{\mathrm{pred}}$ denotes the prediction horizon. When
$T_{\mathrm{pred}}=1$, the task reduces to one-step prediction. In this
work, we focus on the multi-step case with $T_{\mathrm{pred}}>1$. The
future CIR snapshots are predicted jointly, rather than generated by
recursive one-step rollout. This design mitigates error accumulation and
matches practical UWA links more closely.

\begin{table}[!t]
\renewcommand{\arraystretch}{1.25}
\caption{Experimental parameters of the underwater acoustic channel sounding sea trial.}
\label{tab:sea_trial_params}
\centering
\begin{tabular}{|>{\centering\arraybackslash}p{0.55\linewidth}|
                >{\centering\arraybackslash}p{0.30\linewidth}|}
\hline
\textbf{Parameter} & \textbf{Value} \\
\hline
Transmission range & $30~\mathrm{km}$ \\
\hline
Water depth & $50~\mathrm{m}$ \\
\hline
Transducer depth below surface & $30~\mathrm{m}$ \\
\hline
Wind speed & $5$--$11~\mathrm{m/s}$ \\
\hline
Carrier frequency & $3.5~\mathrm{kHz}$ \\
\hline
Symbol rate & $1000~\mathrm{bps}$ \\
\hline
Sampling rate $f_s$ & $40~\mathrm{kHz}$ \\
\hline
Signal duration & $1.5~\mathrm{s}$ \\
\hline
Length of m-sequence & $31$ \\
\hline
Chip duration, $T_c$ & $1~\mathrm{ms}$ \\
\hline
m-sequence period, $T_g$ & $31~\mathrm{ms}$ \\
\hline
\end{tabular}
\end{table}

\section{Dataset Acquisition and Characterization}
\label{sec:dataset}

The measured UWA channel data were obtained from a sea trial conducted in the Taiwan Strait in July 2014. The trial geometry is illustrated in Fig.~\ref{fig:sea_trial}, and the main measurement parameters are summarized in Table~\ref{tab:sea_trial_params}. As illustrated in Fig. \ref{fig:channelSounding}, a dynamic m-sequence-based channel sounding method was employed to capture the temporal evolution of the UWA channel with chip-level resolution. Specifically, circularly shifted m-sequences were transmitted over the acoustic link, and the received signals were processed via sliding correlation with the corresponding local reference sequences. This processing yielded a sequence of time-varying CIR estimates. Unlike conventional m-sequence sounding, which generally provides one CIR estimate per sequence period, the adopted method generates CIR snapshots at intervals of one chip, thereby enabling chip-level characterization of the channel variation. These CIR snapshots were finally organized as the measured UWA channel samples used in this work.

Two measured datasets, denoted by SO-06 and SO-07, are used for evaluation. After channel sounding and matched-correlation processing, each dataset is organized into a CIR matrix $\mathbf{H}\in\mathbb{C}^{N_s\times K}$, where $N_s$ denotes the number of temporal observations and $K$ denotes the number of effective delay taps. Along the time dimension, each dataset consists of $N_s=1500$ CIR snapshots, sampled at an update interval of $T_c=1~\mathrm{ms}$, corresponding to a total observation duration of $N_s T_c=1.5~\mathrm{s}$. Along the delay dimension, the original CIR associated with one m-sequence period covers $T_g=31~\mathrm{ms}$. Given the sampling rate of $f_s=40~\mathrm{kHz}$, this delay window contains $T_g f_s=1240$ waveform samples. These samples are further downsampled by a factor of 10, yielding a compact delay-domain representation with $K=124$ effective taps and a tap spacing of $0.25~\mathrm{ms}$.

\begin{figure}[!t]
\centering
\includegraphics[width=1\columnwidth]{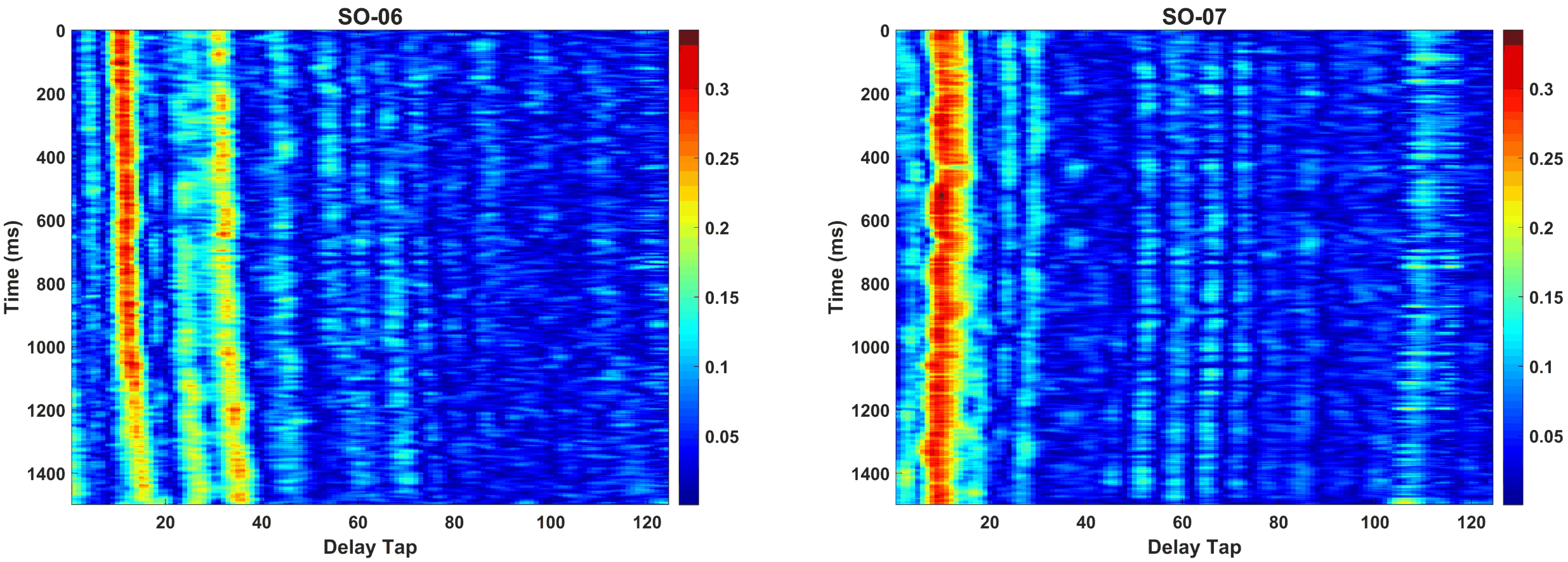}
\caption{Measured time-delay-domain CIRs for SO-06 and SO-07.}
\label{fig:CIR}
\end{figure}

\begin{figure}[!t]
\centering
\includegraphics[width=0.9\columnwidth]{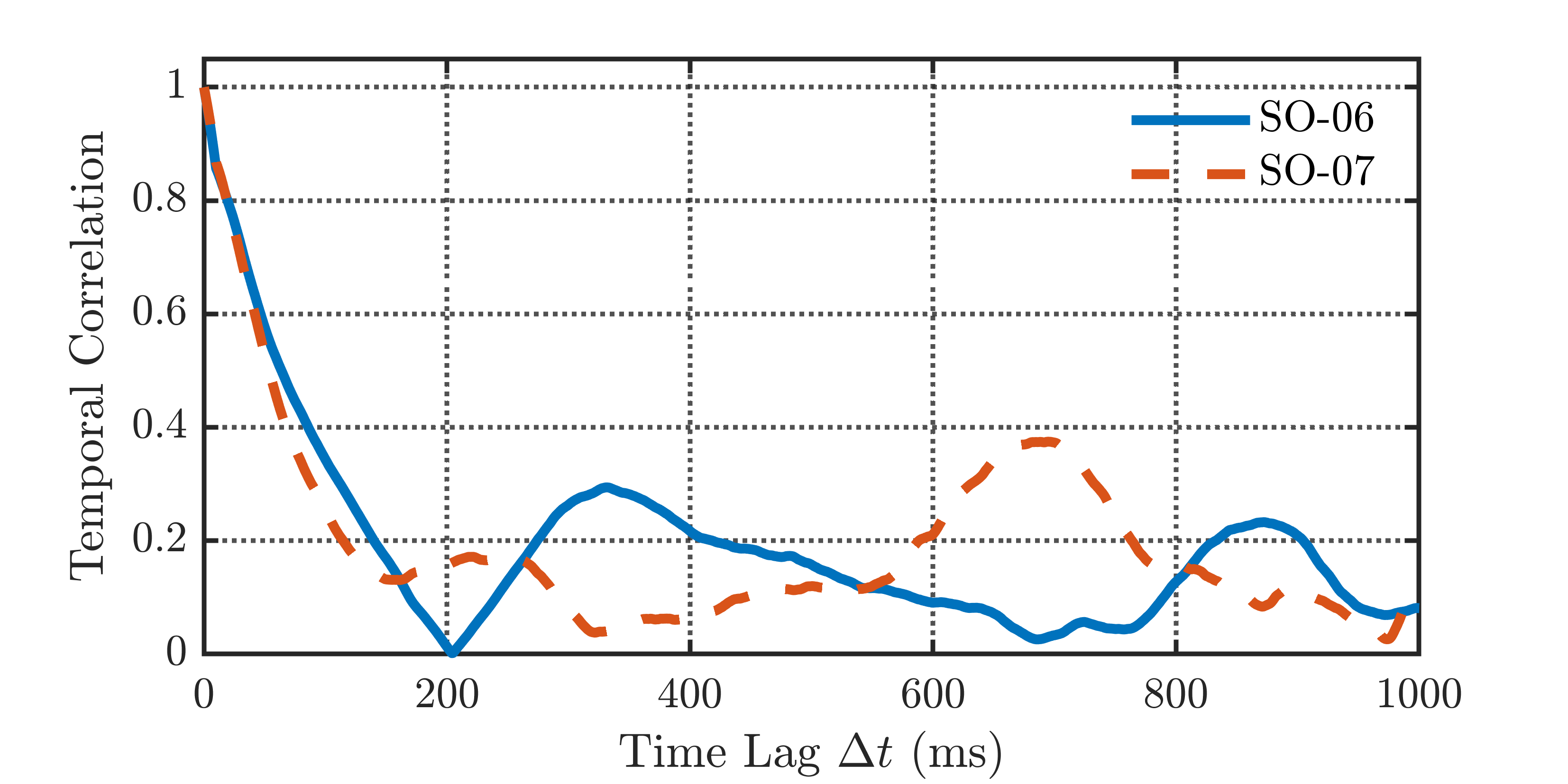}
\caption{Temporal correlation of the measured CIRs for SO-06 and SO-07.}
\label{fig:Temporal_Correlation}
\end{figure}

Fig.~\ref{fig:CIR} shows that the two datasets have distinct delay-domain multipath structures. SO-06 contains more persistent delayed components, whereas in SO-07, most of the channel energy is concentrated around the earliest taps. Both channels also exhibit amplitude fluctuations, slight delay migration, and intermittent path activity, which indicate pronounced time-varying multipath propagation.

To quantify the temporal variation, we compute the normalized temporal correlation of the mean-removed CIR vectors. Let $\mathbf{h}_c[n]=\mathbf{h}[n]-\langle\mathbf{h}[n]\rangle_n \in \mathbb{C}^{K \times 1}$. For temporal lag $\Delta t=\ell T_c$, the correlation is defined as:
\begin{equation}
\rho_t(\Delta t) =
\frac{
\left| \big\langle \left(\mathbf{h}_c[n]\right)^{H}\mathbf{h}_c[n+\ell] \big\rangle_n \right|
}{
\sqrt{
\big\langle \left|\mathbf{h}_c[n]\right|_2^2 \big\rangle_n
\big\langle \left|\mathbf{h}_c[n+\ell]\right|_2^2 \big\rangle_n
}}.
\label{eq:temporal_correlation}
\end{equation}

As shown in Fig.~\ref{fig:Temporal_Correlation}, both datasets exhibit rapid correlation decay at short lags, indicating limited temporal coherence. Their nonmonotonic behavior at longer lags further suggests residual temporal dependence that cannot be fully captured by a simple exponential-decay model.

\begin{figure}[!t]
\centering
\includegraphics[width=1\columnwidth]{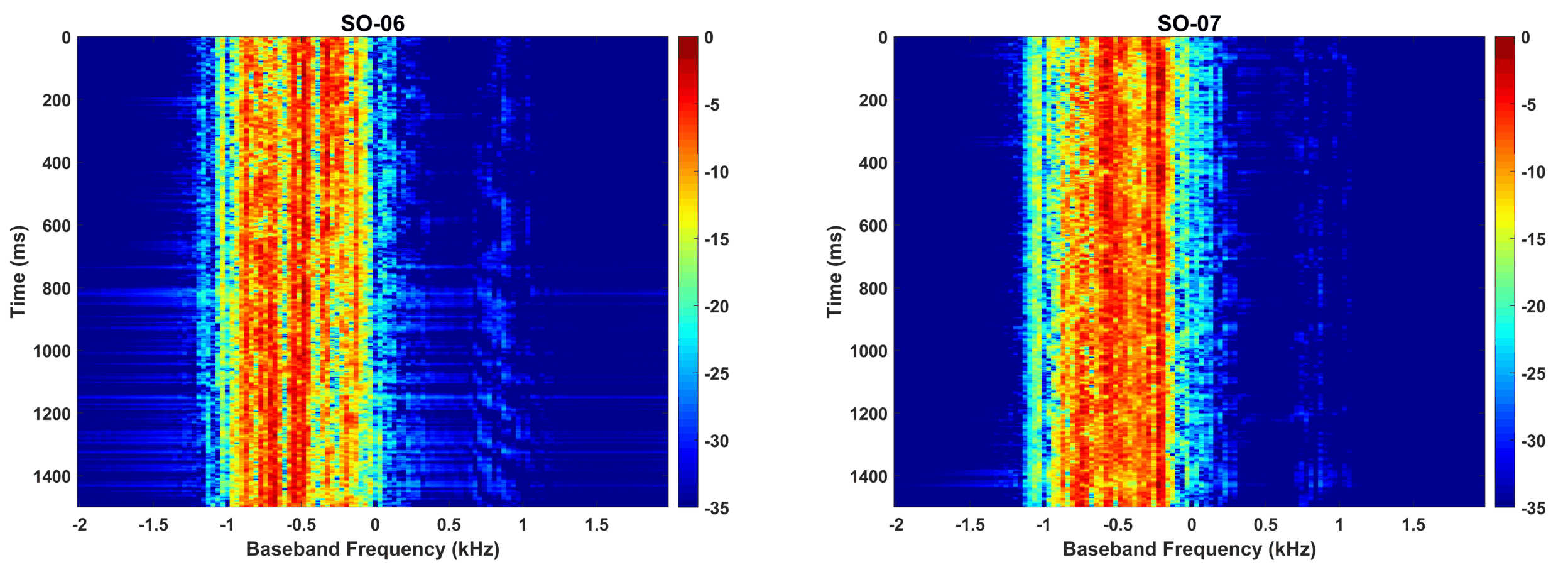}
\caption{Measured time-frequency-domain CFRs for SO-06 and SO-07.}
\label{fig:CFR}
\end{figure}

\begin{figure}[!t]
\centering
\includegraphics[width=0.9\columnwidth]{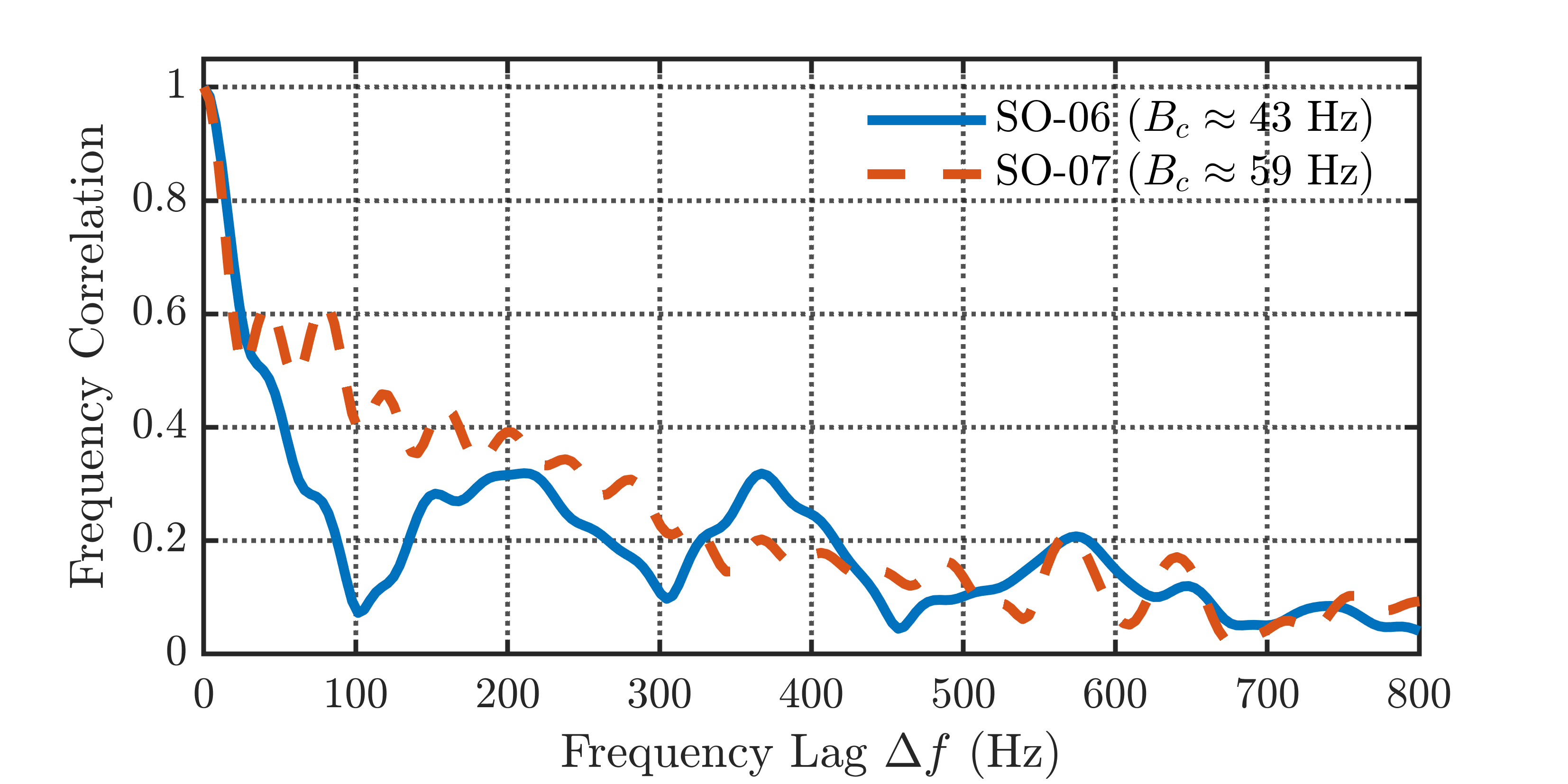}
\caption{Frequency correlation of the measured CFRs for SO-06 and SO-07.}
\label{fig:Frequency_Correlation}
\end{figure}

The corresponding CFRs in Fig.~\ref{fig:CFR} reveal significant in-band frequency selectivity caused by multipath interference. SO-06 exhibits more fragmented spectral patterns and stronger local fluctuations, whereas SO-07 maintains a smoother and more persistent high-energy band.

Frequency selectivity is further characterized by the normalized spaced-frequency correlation. Let $\mathbf{H}_c[f]$ denote the mean-removed CFR vector at frequency bin $f$. For frequency separation $\Delta f$ with bin offset $m$,
\begin{equation}
\rho_f(\Delta f) =
\frac{
\left| \big\langle (\mathbf{H}_c[f])^H \mathbf{H}_c[f+m] \big\rangle_f \right|
}{
\sqrt{
\big\langle |\mathbf{H}_c[f]|_2^2 \big\rangle_f
\big\langle |\mathbf{H}_c[f+m]|_2^2 \big\rangle_f
}}.
\label{eq:freq_correlation}
\end{equation}
Using the threshold definition $B_c=\inf\{\Delta f>0\mid \rho_f(\Delta f)\le 0.5\}$, Fig.~\ref{fig:Frequency_Correlation} gives coherence bandwidths of approximately $43~\mathrm{Hz}$ for SO-06 and $59~\mathrm{Hz}$ for SO-07, confirming stronger frequency selectivity in SO-06.

Overall, SO-06 and SO-07 show distinct delay-energy distributions, temporal correlation patterns, and coherence bandwidths. These propagation-dependent differences increase the difficulty of multi-step CIR prediction, especially when only a limited number of snapshots are available. 
This challenge provides a direct motivation for the subsequent design of our framework.

\section{The Proposed LLM4UACP Framework}
\label{sec:framework}

\subsection{Dual-Domain Feature Extraction and Linear Bypass}

Building on the complementary characteristics analyzed in Section~\ref{sec:dataset}, and inspired by recent dual-domain channel prediction architectures \cite{shi2025channelmamba}, the proposed dual-domain feature extraction module jointly exploits both channel representations. The delay-domain branch extracts delay-time features from the normalized CIR $\mathbf{H}_D \in \mathbb{R}^{2 \times T_{obs} \times K}$, capturing the temporal evolution of resolvable multipath components. In parallel, the frequency-domain branch processes the normalized CFR $\mathbf{H}_F \in \mathbb{R}^{2 \times T_{obs} \times K}$ to explicitly model frequency selectivity and coherence-bandwidth-related spectral correlations. Here, the first dimension denotes the real and imaginary components, $T_{obs}$ denotes the observation length, and $K$ denotes the number of delay or frequency bins.
\begin{figure*}[!t]
\centering
\includegraphics[width=1\textwidth]{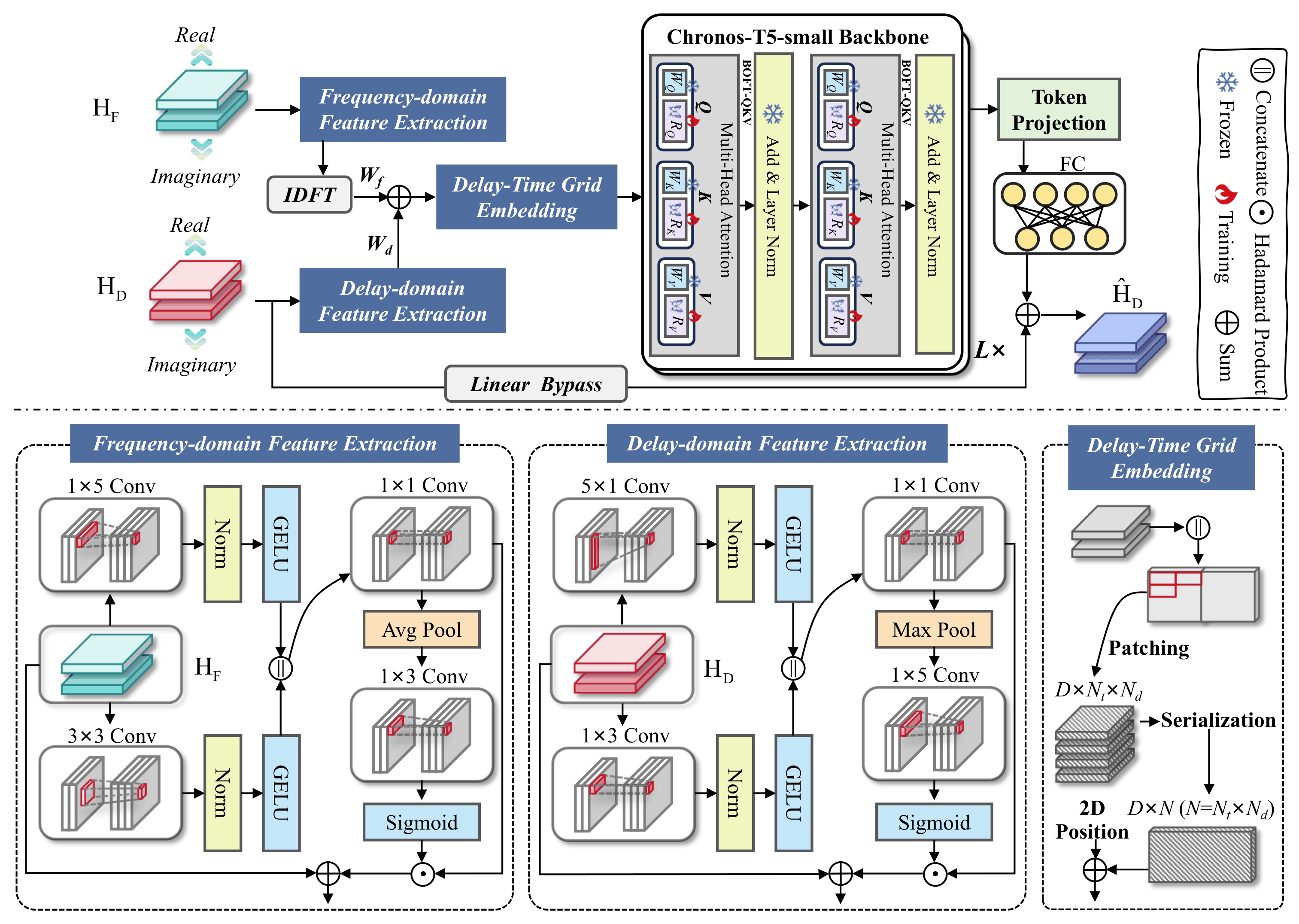}
\caption{Architecture of the proposed LLM4UACP.}
\label{fig:framework}
\end{figure*}

\subsubsection{Linear Bypass}

Over a short prediction interval, the measured channel retains correlation over finite lags. This property allows a portion of the future channel response to be approximated from recent observations through a linear projection in the time dimension. We use a linear bypass from the observed delay-domain sequence to the prediction interval, leaving the nonlinear branch to model the residual nonstationary and nonlinear variations.

Concretely, the delay-domain $\mathbf{H}_D$ is first flattened into $\bar{\mathbf{H}}_D \in \mathbb{R}^{T_{obs} \times 2K}$ and then linearly projected as:
\begin{equation}
\bar{\mathbf{H}}_{skip} = \mathbf{W}_{skip}\bar{\mathbf{H}}_D + \mathbf{B}_{skip},
\label{eq:skip}
\end{equation}
where $\mathbf{W}_{skip} \in \mathbb{R}^{T_{pred} \times T_{obs}}$ is the temporal projection matrix and $\mathbf{B}_{skip} \in \mathbb{R}^{T_{pred} \times 2K}$ is the bias. The output is reshaped into $\mathbf{H}_{skip} \in \mathbb{R}^{2 \times T_{pred} \times K}$ and added to the nonlinear branch output at the channel reconstruction stage.

\subsubsection{Delay-Domain Feature Extraction}

The CIR variations show distinct characteristics along the temporal and delay dimensions. Dominant paths often remain over several consecutive observations, whereas delay-domain dependencies are mainly limited to neighboring taps because of the sparsity of the channel response. This distinction motivates the use of two convolutional branches, with one branch modeling temporal evolution and the other capturing local delay-domain dependencies.

The delay-domain input is processed in the temporal and delay directions as:
\begin{equation}
\mathbf{U}_{time} = g\left( \mathcal{B}\left( \mathbf{W}_{time} \ast \mathbf{H}_D \right) \right),
\label{eq:utime}
\end{equation}
\begin{equation}
\mathbf{U}_{delay} = g\left( \mathcal{B}\left( \mathbf{W}_{delay} \ast \mathbf{H}_D \right) \right),
\label{eq:udelay}
\end{equation}
where $\mathbf{W}_{time} \in \mathbb{R}^{16 \times 2 \times 5 \times 1}$ and $\mathbf{W}_{delay} \in \mathbb{R}^{16 \times 2 \times 1 \times 3}$ are the convolution kernels of the time and delay branches, respectively. The $5 \times 1$ kernel captures the evolution of each delay tap across consecutive observations, while the $1 \times 3$ kernel aggregates information from neighboring taps. $\mathcal{B}(\cdot)$ and $g(\cdot)$ denote batch normalization and the GELU activation function, respectively.

The two directional features are concatenated along the channel dimension and fused through a $1 \times 1$ convolution:
\begin{equation}
\mathbf{U}_D = \mathbf{W}_{fuse}^{(D)} \ast \left[ \mathbf{U}_{time} \parallel \mathbf{U}_{delay} \right],
\label{eq:ud_fuse}
\end{equation}
where $\mathbf{W}_{fuse}^{(D)} \in \mathbb{R}^{2 \times 32 \times 1 \times 1}$.

A delay-domain attention map is subsequently employed to enhance prominent multipath components. Since dominant paths may be strongly activated only at discrete time instants, max pooling is applied along the time dimension while preserving the time dimension as a singleton, 
\begin{equation}
\mathbf{A}_D = \sigma \left( \mathbf{W}_{attn}^{(D)} \ast \max_{t} \mathbf{U}_D \right).
\label{eq:ad}
\end{equation}
Here, $\max_t \mathbf{U}_D \in \mathbb{R}^{2 \times 1 \times K}$ denotes the temporally pooled feature, and $\mathbf{W}_{attn}^{(D)} \in \mathbb{R}^{2 \times 2 \times 1 \times 5}$ denotes the convolutional kernel with two input channels, two output channels, and a kernel size of $1 \times 5$. This convolution is performed on the temporally pooled feature along the delay dimension. The resulting attention map $\mathbf{A}_D \in \mathbb{R}^{2 \times 1 \times K}$ is then broadcast along the temporal dimension and applied to $\mathbf{U}_D$.

The delay-domain output is obtained through residual enhancement:
\begin{equation}
\mathbf{Z}_D = \mathbf{H}_D + \mathbf{U}_D \odot \mathbf{A}_D,
\label{eq:zd_final}
\end{equation}
where $\odot$ denotes element-wise multiplication. The residual path preserves the original CIR structure while the attention-weighted term strengthens locally dominant multipath responses.

\subsubsection{Frequency-Domain Feature Extraction}

The preceding frequency-correlation analysis shows that the measured CFRs have limited coherence bandwidth and strong frequency selectivity. The frequency-domain feature extraction module is therefore designed to capture local spectral correlation and short-range CFR variations across consecutive time snapshots.

The frequency-domain input is processed through two branches as:
\begin{equation}
\mathbf{U}_{freq} = g\left( \mathcal{B}\left( \mathbf{W}_{freq} \ast \mathbf{H}_F \right) \right),
\label{eq:ufreq}
\end{equation}
\begin{equation}
\mathbf{U}_{grid} = g\left( \mathcal{B}\left( \mathbf{W}_{grid} \ast \mathbf{H}_F \right) \right),
\label{eq:ugrid}
\end{equation}
where $\mathbf{W}_{freq} \in \mathbb{R}^{16 \times 2 \times 1 \times 5}$ and $\mathbf{W}_{grid} \in \mathbb{R}^{16 \times 2 \times 3 \times 3}$ denote the convolution kernels of the frequency and local grid branches, respectively. The $1 \times 5$ kernel extracts local spectral patterns across neighboring frequency bins, whereas the $3 \times 3$ kernel jointly aggregates consecutive time snapshots and neighboring frequency bins to capture local variations in the time-frequency grid.

The outputs of the two branches are concatenated along the channel dimension and fused using a $1 \times 1$ convolution
\begin{equation}
\mathbf{U}_F = \mathbf{W}_{fuse}^{(F)} \ast \left[ \mathbf{U}_{freq} \parallel \mathbf{U}_{grid} \right],
\label{eq:ufuse}
\end{equation}
where $\mathbf{W}_{fuse}^{(F)} \in \mathbb{R}^{2 \times 32 \times 1 \times 1}$ denotes the fusion kernel. The frequency branch captures local spectral correlations, while the local time-frequency branch captures short-range joint variations across time and frequency.

A frequency-domain attention map is generated from the fused feature to emphasize stable spectral responses. Because the CFR is locally correlated and relatively smooth along the frequency axis, temporal average pooling is used to suppress transient fluctuations and obtain a stable spectral representation:
\begin{equation}
\mathbf{A}_F = \sigma\left( \mathbf{W}_{attn}^{(F)} \ast \mathbb{E}_t\left[\mathbf{U}_F\right] \right),
\label{eq:af}
\end{equation}
where $\mathbf{W}_{attn}^{(F)} \in \mathbb{R}^{2 \times 2 \times 1 \times 3}$ is the convolution kernel used to generate the attention map.

The frequency-domain output is computed using a residual connection
\begin{equation}
\mathbf{Z}_F = \mathbf{H}_F + \mathbf{U}_F \odot \mathbf{A}_F.
\label{eq:zf_final}
\end{equation}
This residual formulation preserves the original CFR structure while adaptively enhancing locally correlated and frequency-selective responses.

\subsubsection{Cross-Domain Feature Fusion}
\label{subsubsec:fusion}

The delay- and frequency-domain representations provide complementary information about the same UWA channel, but their relative contributions vary with channel conditions. Fixed fusion methods, such as averaging and predefined weighted summation, use the same weights for all samples and cannot adapt to varying channel conditions. If one domain is less informative under a particular channel condition, fixed fusion may weaken useful features from the other domain or place excessive weight on unreliable information. We therefore first map the frequency-domain representation onto the delay-domain grid using a fixed IDFT and then apply a lightweight sample-level gate to adaptively balance the two representations.

The delay-domain feature $\mathbf{Z}_D$ and frequency-domain feature $\mathbf{Z}_F$ are derived from the same UWA channel but are indexed in different domains. Before fusion, $\mathbf{Z}_F$ is mapped onto a delay-indexed grid by a fixed orthonormal inverse DFT basis along the $K$ dimension. Specifically, the two channels of $\mathbf{Z}_F$ are first combined into a complex-valued latent feature,
\begin{equation}
\mathbf{Z}_F^{c}=\mathbf{Z}_F^{(1)}+j\mathbf{Z}_F^{(2)} ,
\label{eq:complex_zf}
\end{equation}
and then transformed as:
\begin{equation}
\tilde{\mathbf{Z}}_F^{c}=\mathbf{Z}_F^{c}\mathbf{F}_{K}^{H},
\label{eq:idft_zf}
\end{equation}
where $\mathbf{F}_{K}^{H}$ is the inverse orthonormal DFT matrix along the $K$ dimension. The aligned feature is represented by its real and imaginary parts as
\begin{equation}
\tilde{\mathbf{Z}}_F =
\left[
\Re\left(\tilde{\mathbf{Z}}_F^{c}\right),
\Im\left(\tilde{\mathbf{Z}}_F^{c}\right)
\right],
\label{eq:aligned_zf}
\end{equation}
where $\tilde{\mathbf{Z}}_F \in \mathbb{R}^{2 \times T_{\mathrm{obs}} \times K}$ matches the dimensionality of $\mathbf{Z}_D$. This parameter-free feature-level Fourier alignment is not intended to perform strict physical CIR reconstruction. Instead, it provides a domain-consistent representation for subsequent feature fusion.

With the frequency-domain feature aligned to the delay-indexed grid, a lightweight sample-level gate adaptively weights the delay-domain and projected frequency-domain features. Each feature is summarized by averaging over time and delay while preserving the real and imaginary channels.
\begin{equation}
\mathbb{E}_s[\mathbf{Z}]
=
\frac{1}{T_{obs}K}
\sum_{t=1}^{T_{obs}}
\sum_{k=1}^{K}
\mathbf{Z}_{:,t,k}
\in \mathbb{R}^{2 \times 1}.
\end{equation}
The descriptors $\mathbb{E}_s[\mathbf{Z}_D]$ and $\mathbb{E}_s[\tilde{\mathbf{Z}}_F]$ are concatenated as $\mathbf{q} \in \mathbb{R}^{4 \times 1}$, and the fusion weights are computed as
\begin{equation}
\boldsymbol{\omega}
=
\mathrm{Softmax}
\left(
\mathbf{W}_{gate}\mathbf{q}
+
\mathbf{b}_{gate}
\right),
\label{eq:gate_fusion_weight}
\end{equation}
where $\boldsymbol{\omega}=[\omega_d,\omega_f]^\top$ gives the sample-level weights of the delay-domain and projected frequency-domain features, respectively.

The fused representation is computed as:
\begin{equation}
\mathbf{H}_{fused}
=
\omega_d \mathbf{Z}_D
+
\omega_f \tilde{\mathbf{Z}}_F.
\label{eq:gate_fused_final}
\end{equation}
The resulting feature $\mathbf{H}_{fused} \in \mathbb{R}^{2 \times T_{obs} \times K}$ preserves the delay-time grid structure and is fed into the subsequent sequence modeling module.

\subsection{Delay-Time Grid Embedding}

To adapt the two-dimensional fused representation to the one-dimensional token sequence required by the pretrained LLM, LLM4UACP maps $\mathbf{H}_{fused}$ to delay-time tokens using patch-based embedding. Each token covers a local neighborhood across consecutive observations and adjacent delay taps, retaining short-range multipath evolution while keeping the sequence length manageable for the backbone.

Specifically, the channel and delay dimensions of $\mathbf{H}_{fused}$ are first merged into an extended delay-related grid
$\mathbf{\tilde{H}}_{fused} \in \mathbb{R}^{T_{obs} \times 2K}$, which is then divided into nonoverlapping patches of size $P_t \times P_d$. Each patch is mapped to a $D$-dimensional embedding through a convolutional projection,
\begin{equation}
\mathbf{E}_{grid} = \mathbf{W}_{emb} \ast_{(P_t,P_d)} \mathbf{\tilde{H}}_{fused} + \mathbf{b}_{emb},
\label{eq:patch_embedding}
\end{equation}
where $\ast_{(P_t,P_d)}$ denotes convolution with kernel size $P_t \times P_d$ and stride $(P_t,P_d)$. Here, $\mathbf{W}_{emb} \in \mathbb{R}^{D  \times P_t \times P_d}$ is the embedding kernel, and $\mathbf{b}_{emb} \in \mathbb{R}^{D}$ is the bias. This operation is equivalent to a shared linear projection applied to vectorized local delay-time neighborhoods. The resulting feature map satisfies
$\mathbf{E}_{grid} \in \mathbb{R}^{D \times N_t \times N_d}$, where $N_t = T_{obs}/P_t$ and $N_d = 2K/P_d$.

However, unfolding the embedded grid into a one-dimensional sequence weakens the explicit correspondence between each token and its original delay-time coordinate. To retain this coordinate information, we add a parameter-free two-dimensional sinusoidal positional encoding and define the final token sequence as:
\begin{equation}
\mathbf{Z}_{token} = \mathcal{M}\left( \mathbf{E}_{grid} \right) + \mathbf{E}_{pos}^{2D},
\label{eq:token_serialization}
\end{equation}
where $\mathcal{M}:\mathbb{R}^{D \times N_t \times N_d} \rightarrow \mathbb{R}^{N \times D}$ denotes the unfolding operation over patch positions, $N = N_t N_d$, $\mathbf{E}_{pos}^{2D} \in \mathbb{R}^{N \times D}$ is stacked in the same order as $\mathcal{M}(\cdot)$, and $\mathbf{Z}_{token} \in \mathbb{R}^{N \times D}$.

\subsection{Domain Adaptation via BOFT}
\label{subsec:speft}

After delay-time grid embedding, the UWA channel observations are represented as continuous channel tokens and fed into a pretrained Transformer backbone. The backbone is the encoder of Chronos-T5-small \cite{ansari2024chronos-T5-small}, a model pretrained for time-series modeling that provides a suitable sequence prior for numerical temporal patterns. Compared with general-purpose LLM backbones, Chronos-T5-small better matches time-series inputs, and its compact scale supports adaptation to UWA channel prediction with limited training data.

Although $\mathbf{Z}_{token}$ has the format required by the backbone, it still encodes channel behavior shaped by underwater multipath propagation, Doppler variation, and frequency-selective fading. With limited UWA data, fine-tuning the full backbone can be inefficient and may lead to overfitting. We therefore adopt Butterfly Orthogonal Fine-Tuning (BOFT) \cite{boft} as a lightweight domain adaptation module while keeping the original backbone parameters frozen. BOFT applies trainable orthogonal transformations to the attention projection space, so the attention module can adjust correlations among delay-time channel tokens while preserving the geometry of the pretrained representation space.

For each attention projection matrix $\mathbf{W}_{m}$ with $m \in \{Q,K,V\}$, BOFT applies a trainable orthogonal transformation $\mathbf{R}_{m}$ to obtain the effective projection matrix
\begin{equation}
\mathbf{W}'_{m}=\mathbf{R}_{m}\mathbf{W}_{m},\quad
m \in \{Q,K,V\},
\label{eq:boft_weight_update}
\end{equation}
where $\mathbf{W}_{m}$ is the frozen pretrained projection matrix and $\mathbf{R}_{m}$ is the trainable orthogonal transformation for domain adaptation. Since $\mathbf{R}_{m}$ is orthogonal, it rotates the projection space while preserving the norm and relative geometry of the pretrained representations.

To reduce the number of trainable parameters, BOFT represents $\mathbf{R}_{m}$ using butterfly-structured factors,
\begin{equation}
\mathbf{R}_{m}=\prod_{i=1}^{L_b}\mathbf{B}_{m,i},
\label{eq:boft_butterfly_factorization}
\end{equation}
where $\mathbf{B}_{m,i}$ denotes the $i$-th butterfly factor and $L_b$ is the number of butterfly factorization stages. This structured parameterization provides a compact form for adapting the query, key, and value projections, which determine attention scores and token aggregation patterns.

With this parameterization, domain adaptation is performed by updating only the BOFT parameters associated with $\mathbf{R}_{Q}$, $\mathbf{R}_{K}$, and $\mathbf{R}_{V}$, while all pretrained Transformer parameters remain frozen. This allows the model to adapt the attention-based interactions among $\mathbf{Z}_{token}$ while retaining the temporal modeling prior of the pretrained backbone. Let $\Theta_{\mathrm{frozen}}$ collect all frozen parameters of the pretrained Transformer, and let $\Theta_{\mathrm{BOFT}}$ denote the trainable BOFT parameters inserted into the query, key, and value projections. Given the token sequence $\mathbf{Z}_{\mathrm{token}} \in \mathbb{R}^{N \times D}$, the backbone output is written as:
\begin{equation}
\mathbf{H}^{(L)}=\mathcal{F}_{\mathrm{LLM}}\left(\mathbf{Z}_{\mathrm{token}};\Theta_{\mathrm{frozen}},\Theta_{\mathrm{BOFT}}\right),
\label{eq:llm_forward}
\end{equation}
where $\mathbf{H}^{(L)} \in \mathbb{R}^{N \times D}$ denotes the final-layer latent token representation.

\subsection{Channel Reconstruction}
\label{sec:reconstruction}

The backbone output $\mathbf{H}^{(L)}$ remains a serialized patch sequence. Channel reconstruction maps this sequence back to the delay-domain channel representation by reversing the tokenization structure, restoring the target prediction steps, adding the linear bypass, and applying inverse normalization.

The inverse unfolding operator $\mathcal{M}^{-1}(\cdot)$ first restores the serialized tokens to the patch grid, and each restored patch token is projected to a local delay segment and concatenated along the delay dimension,
\begin{equation}
\bar{\mathbf{H}}_{delay}
= \mathcal{F}\left(
\mathcal{M}^{-1}\left( \mathbf{H}^{(L)} \right)\mathbf{W}_{rec}
+ \mathbf{b}_{delay}
\right)
\in \mathbb{R}^{N_t \times 2K}.
\label{eq:decoder_output}
\end{equation}
where $\mathcal{M}^{-1}\left( \mathbf{H}^{(L)} \right) \in \mathbb{R}^{N_t \times N_d \times D}$, $\mathbf{W}_{rec} \in \mathbb{R}^{D \times P_d}$, and $\mathbf{b}_{delay} \in \mathbb{R}^{P_d}$. The operator $\mathcal{F}(\cdot)$ reshapes the projected tokens from $\mathbb{R}^{N_t \times N_d \times P_d}$ to $\mathbb{R}^{N_t \times 2K}$ by concatenating the $N_d$ delay segments of length $P_d$ in their original order for each time-patch index, with $N_d P_d = 2K$.

The recovered delay representation is mapped from the reduced time resolution $N_t$ to the prediction horizon $T_{pred}$ by
\begin{equation}
\bar{\mathbf{H}}_{base} = \mathbf{W}_{pred} \bar{\mathbf{H}}_{delay} + \mathbf{B}_{pred} \in \mathbb{R}^{T_{pred} \times 2K},
\label{eq:final_prediction}
\end{equation}
where $\mathbf{W}_{pred} \in \mathbb{R}^{T_{pred} \times N_t}$ and $\mathbf{B}_{pred} \in \mathbb{R}^{T_{pred} \times 2K}$.

The output is reshaped into the two-channel form $\mathbf{H}_{base} \in \mathbb{R}^{2 \times T_{pred} \times K}$ and combined with the linear bypass prediction,
\begin{equation}
\hat{\mathbf{H}}_{norm} = \mathbf{H}_{base} + \mathbf{H}_{skip},
\label{eq:skip_addition}
\end{equation}
where $\hat{\mathbf{H}}_{norm}$ remains in the globally normalized delay-domain scale.

Finally, inverse global Z-score normalization recovers the complex-valued CIR. During preprocessing, the scalar mean $\mu_D^{(c)}$ and standard deviation $\sigma_D^{(c)}$ are computed from all training entries of the delay-domain component $c$, where $c=1$ and $c=2$ correspond to the real and imaginary channels, respectively. The complex-valued prediction $\hat{\mathbf{H}}^{D}_{pred} \in \mathbb{C}^{T_{pred} \times K}$ is recovered as:
\begin{equation}
\hat{\mathbf{H}}^{D}_{pred}=\sum_{c=1}^{2}\left[\hat{\mathbf{H}}_{norm}^{(c)}\left(\sigma_D^{(c)}+\epsilon\right)+\mu_D^{(c)}\right]j^{\,c-1},
\label{eq:final_physical_reconstruct}
\end{equation}
where $\hat{\mathbf{H}}_{norm}^{(c)} \in \mathbb{R}^{T_{pred}\times K}$ denotes the $c$-th normalized output channel, $\epsilon$ is a small constant for numerical stability, and $j$ denotes the imaginary unit.

\section{Model Training and Evaluation}
\label{sec:training}

\subsection{Data Preprocessing and Sample Construction}
\label{subsec:preprocessing}

Let $\tilde{\mathbf{H}}_{D}\in\mathbb{C}^{N_{\mathrm{s}}\times K}$ denote the collected delay-domain CIR sequence. Its frequency-domain representation $\tilde{\mathbf{H}}_{F}$ is obtained by applying the $K$-point orthonormal Fourier transform along the delay dimension. Since the transform is applied independently to each temporal snapshot, the delay-domain and frequency-domain sequences remain strictly time-aligned.

We synchronously split the paired sequences $\tilde{\mathbf{H}}_{D}$ and $\tilde{\mathbf{H}}_{F}$ along the temporal axis into nonoverlapping training, validation, and testing subsets with ratios of 70\%, 10\%, and 20\%, respectively. This split precedes normalization and sample construction to prevent information leakage. Both domains share identical temporal indices, thereby preserving the one-to-one correspondence between delay-domain samples and frequency-domain samples.

For each domain $x\in\{D,F\}$, the complex-valued channel sequence is decomposed into real and imaginary components, denoted by $\tilde{\mathbf{H}}_{x}^{(1)}=\Re(\tilde{\mathbf{H}}_{x})$ and $\tilde{\mathbf{H}}_{x}^{(2)}=\Im(\tilde{\mathbf{H}}_{x})$, respectively. Global z-score normalization is then applied separately to each domain and each component. This normalization stabilizes the numerical scale while preserving the relative energy distribution of the channel coefficients.

After normalization, input-output samples are generated within each subset using a sliding window with stride $s=1$. Each sample contains $T_{obs}$ historical frames as the input and the following $T_{pred}$ frames as the prediction target. The sliding windows are constructed only within the same subset and do not cross the training-validation or validation-testing boundary.

\begin{table}[!t]
\renewcommand{\arraystretch}{1.5}
\setlength{\tabcolsep}{3.5pt}
\caption{Hyperparameter configuration of LLM4UACP.}
\label{tab:hyperparameters}
\centering
\scriptsize
\begin{tabular}{|>{\centering\arraybackslash}p{0.50\columnwidth}|
                >{\centering\arraybackslash}p{0.38\columnwidth}|}
\hline
\textbf{Hyperparameter} & \textbf{Value} \\
\hline
 $T_{\mathrm{obs}},T_{\mathrm{pred}}$ 
& $16,4$ \\
\hline
Delay taps / patch size, $K$, $(P_t,P_d)$ 
& $124$, $(4,8)$ \\
\hline
Hidden dimension / layers / heads 
& $512,6,8$ \\
\hline
BOFT block size / butterfly stages 
& $8,4$ \\
\hline
Optimizer / batch size, $B$ 
& AdamW, $128$ \\
\hline
Learning rate / weight decay, $(\eta,\lambda)$ 
& $(2\times10^{-4},\,1\times10^{-4})$ \\
\hline
LR scheduler / decay factor / patience 
& Plateau, $0.5$, $5$ \\
\hline
Maximum epochs / early-stopping patience 
& $500$, $10$ \\
\hline
\end{tabular}
\end{table}

\subsection{Model Configuration and Parameter Initialization}
\label{subsec:config_init}

Table~\ref{tab:hyperparameters} summarizes the main hyperparameter settings, including delay-time patch construction, architectural dimensions, and the optimization setup. To improve training stability, the newly introduced modules are initialized according to layer type. Convolutional kernels use Kaiming Normal initialization to maintain activation variance under the GELU nonlinearity, whereas the weight matrices in the linear bypass and channel reconstruction layers use Xavier Uniform initialization. All bias terms are initialized to zero.

\subsection{Loss Function}
\label{subsec-loss-function}

Accurate multi-step channel prediction requires reliable reconstruction of each future CIR snapshot and consistent temporal evolution across the predicted sequence. We therefore combine an NMSE reconstruction term, which measures the relative error of the complex-valued CIRs, with a first-order temporal-gradient consistency term that matches the frame-to-frame variations of the predicted and measured CIR sequences.

The predicted channel and corresponding ground truth over the prediction steps are denoted by $\hat{\mathbf{H}}, \mathbf{H} \in \mathbb{R}^{2 \times T_{pred}\times K}$, respectively. The reconstruction term is defined as: 
\begin{equation}
\mathcal{L}_{\mathrm{nmse}}=\mathbb{E}\left[\frac{\left\|\hat{\mathbf{H}}-\mathbf{H}\right\|_F^2}{\left\|\mathbf{H}\right\|_F^2+\epsilon}\right],
\label{eq-loss-nmse}
\end{equation}
where $\epsilon$ is a small positive constant used to prevent numerical instability. The size-two dimension of $\hat{\mathbf{H}}$ and $\mathbf{H}$ represents the real and imaginary components of the complex channel. This real-valued representation therefore has the same squared Frobenius norm as the corresponding complex-valued channel, which makes the training NMSE consistent with the complex-domain NMSE used for evaluation.

The first-order gradient loss is defined as:
\begin{equation}
\mathcal{L}_{\mathrm{grad}}=\mathbb{E}\left[\frac{1}{\Omega}\left\|\nabla_t \hat{\mathbf{H}}-\nabla_t \mathbf{H}\right\|_F^2\right],
\label{eq-loss-grad}
\end{equation}
where $\nabla_t$ denotes the first-order forward-difference operator along the prediction steps, and $\Omega=(T_{pred}-1)\cdot 2K$ normalizes over all difference entries.

The two loss terms are complementary but can have different numerical scales. These scales may vary with propagation conditions, mobility, and Doppler spread. We therefore use homoscedastic uncertainty weighting \cite{loss2} instead of manually tuning a fixed scalar coefficient. Two learnable log-variance parameters, $s_n$ and $s_g$, are assigned to the NMSE and gradient losses, respectively. The total training objective is given by
\begin{equation} 
\mathcal{L}_{\mathrm{tot}}
=
\sum_{x\in\{n,g\}}
\left(
\exp(-s_x)\mathcal{L}_x+s_x
\right),
\label{eq-loss-total}
\end{equation}
where $\mathcal{L}_n=\mathcal{L}_{\mathrm{nmse}}$ and $\mathcal{L}_g=\mathcal{L}_{\mathrm{grad}}$. The parameters $s_n$ and $s_g$ are initialized to zero and optimized jointly with the network parameters, and $\exp(-s_x)$ acts as the effective weight of the corresponding loss term.

\subsection{Performance Evaluation Metrics}
\label{subsec-metrics}

To evaluate the proposed LLM4UACP framework, we consider both physical-layer channel prediction accuracy and the reliability of the downstream communication system. The Normalized Mean Squared Error (NMSE) and Bit Error Rate (BER) are used as the primary evaluation metrics.

\subsubsection{Normalized Mean Squared Error (NMSE)}
We use NMSE to measure the reconstruction error on the testing set and evaluate the prediction accuracy of the proposed LLM4UACP. Training uses decoupled real-valued representations, whereas evaluation uses reconstructed complex-valued CIRs to assess channel prediction performance in the complex domain. For the testing set $\mathcal{D}_{\mathrm{test}}$ with $N_{\mathrm{test}}$ samples, the average NMSE is defined as:
\begin{equation}
\mathrm{NMSE} = \frac{\sum_{i=1}^{N_{\mathrm{test}}} \left\| \hat{\mathbf{H}}_{\mathrm{pred}}^{(i)} - \mathbf{H}_{\mathrm{target}}^{(i)} \right\|_F^2}{\sum_{i=1}^{N_{\mathrm{test}}} \left\| \mathbf{H}_{\mathrm{target}}^{(i)} \right\|_F^2},
\label{eq-nmse-test}
\end{equation}
where $\hat{\mathbf{H}}_{\mathrm{pred}}^{(i)}$ and $\mathbf{H}_{\mathrm{target}}^{(i)} \in \mathbb{C}^{T_{\mathrm{pred}} \times K}$ denote the predicted complex-valued channel and the corresponding ground-truth channel for the $i$-th sample in the testing set, respectively. This metric reflects how well the model predicts the future evolution of nonstationary UWA channels over the prediction steps. A lower NMSE indicates higher prediction accuracy, leading to more accurate multipath tracking and more effective compensation for Doppler-induced channel variations.

\subsubsection{Bit Error Rate (BER)}

BER measures downstream communication reliability when detection uses the predicted channel:
\begin{equation}
    \mathrm{BER} = \frac{N_{\mathrm{err}}}{N_{\mathrm{bit}}},
\end{equation}
where $N_{\mathrm{err}}$ and $N_{\mathrm{bit}}$ denote the numbers of erroneous and transmitted bits, respectively.

\section{Experimental Results and Discussion}
\label{sec:expreiment}

All experiments were run on Ubuntu 22.04.5 LTS with an Intel Core i9-13900K CPU, an NVIDIA GeForce RTX 4080 GPU with 16 GB memory, NVIDIA driver 535.288.01, and CUDA 12.2.

\subsection{Baseline Models for Comparison}
Most existing baselines target one-step prediction. For fair comparison in our multi-step setting, we implement multi-step prediction via recursive rollout, feeding each predicted CIR back as the next-step input. For RLS, standard recursion updates linear prediction coefficients online using the inverse correlation matrix; each delay tap is predicted independently with filter order $M_{order}=16$, prediction step $T_{pred}=4$, forgetting factor $\lambda=0.75$, and inverse correlation matrix initialized as $\delta^{-1}\mathbf{I}$ with $\delta=0.01$, while real and imaginary components are processed separately for complex-domain prediction. For CsiPreNet~\cite{csiprenet}, we keep its CNN-LSTM paradigm for joint frequency-temporal modeling, where two 1-D convolutional layers with kernel size 5 and max-pooling extract local frequency features, and a two-layer bidirectional LSTM with 256 hidden units models temporal dynamics. For AttLSTMPreNet-Type2~\cite{attlstm}, we keep its attention-LSTM framework, where soft attention reweights the input sequence before a two-layer bidirectional LSTM with hidden size 256, followed by dropout and a regression head for channel prediction. For STL-Transformer, we adopt a standard encoder-decoder Transformer with $d_{model}=64$ and 4 attention heads, with remaining settings following~\cite{multi-task} unless otherwise specified.

\begin{table*}[!t]
\caption{Overall Evaluation Results of LLM4UACP and Baseline Models}
\label{tab:overall_evaluation_results}
\centering
\footnotesize
\setlength{\tabcolsep}{6.3pt}
\renewcommand{\arraystretch}{1.50}
\begin{tabular}{|c|cc|ccc|cc|}
\hline
\multirow{2}{*}{\textbf{Model}}
& \multicolumn{2}{c|}{\textbf{NMSE}}
& \multicolumn{3}{c|}{\textbf{Model Complexity}}
& \multicolumn{2}{c|}{\textbf{Inference Efficiency}} \\
\cline{2-8}

& \textbf{\shortstack{SO-06}}& \textbf{\shortstack{SO-07}}& \textbf{\shortstack{Total\\Params (M)}}& \textbf{\shortstack{Trainable\\Params (M)}}& \textbf{\shortstack{FLOPs (G)}}& \textbf{\shortstack{Avg.\\Latency (ms)}}& \textbf{\shortstack{Std.\\Latency (ms)}} \\
\hline

\textbf{RLS}
& 0.3375
& 0.2989
& -
& -
& -
& 0.786
& 0.206 \\
\hline

\textbf{AttLSTMPreNet}
& 0.6065
& 0.3765
& 3.13& 3.13 & 0.102
& 1.130& 0.288\\
\hline

\textbf{CsiPreNet}
& 0.6397
& 0.4868
& 11.03
& 11.03
& 1.649
& 2.068& 0.395\\
\hline

\textbf{STL-Transformer}
& 0.3586
& 0.2688
& 0.15
& 0.15
& 0.041
& 1.706& 0.507\\
\hline

\textbf{LLM4UACP}
& \textbf{0.0411}& \textbf{0.0399}
& 21.08
&  0.11
& 12.103
&  7.602& 0.097 \\
\hline

\end{tabular}
\end{table*}

\begin{figure}[!t]
\centering
\includegraphics[width=1\columnwidth]{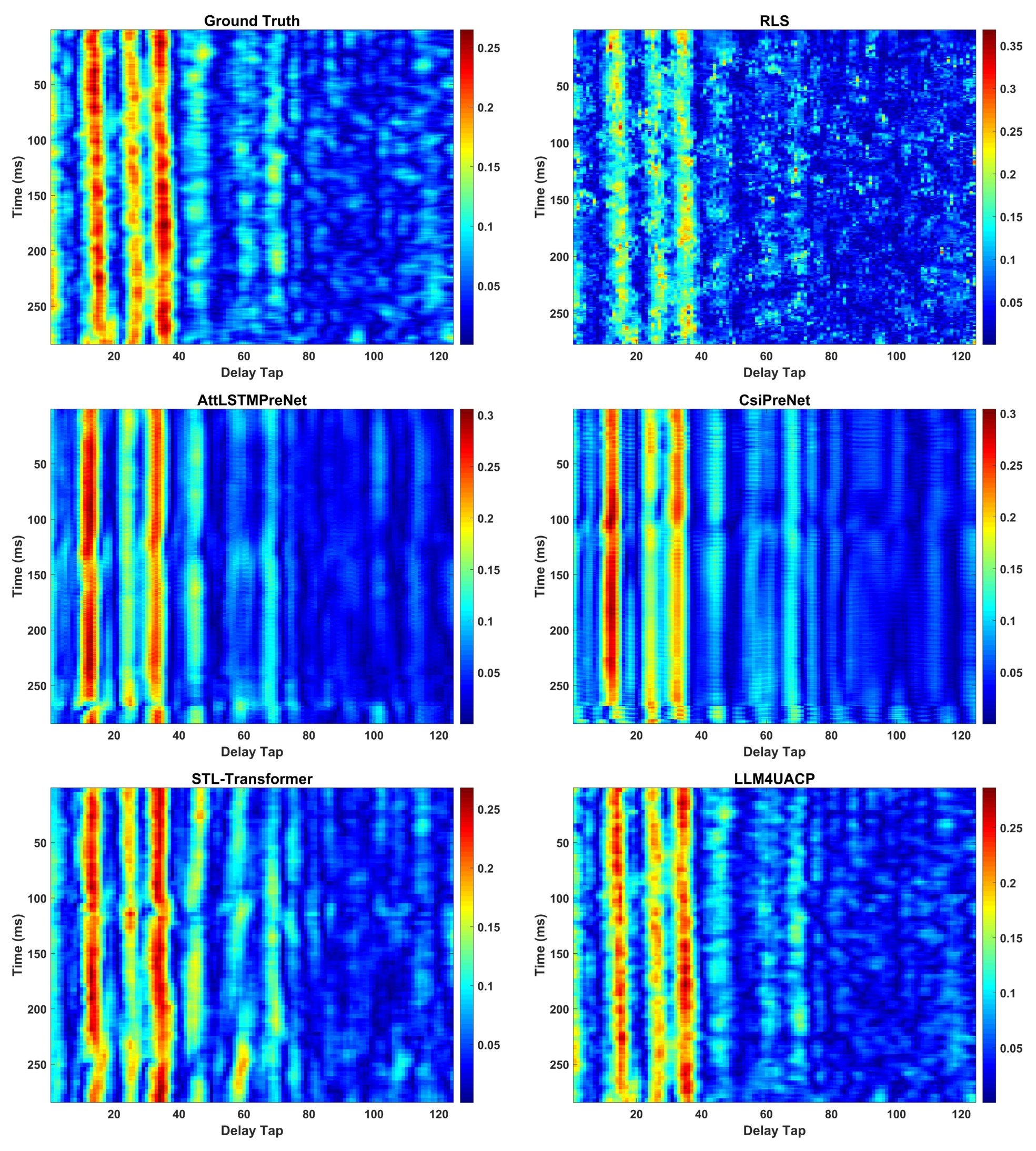}
\caption{Delay-time CIR prediction results on the SO-06 dataset.}
\label{fig:so06_prediction}
\end{figure}

\subsection{Overall Prediction Performance}
\label{subsec:overall_performance}

Table~\ref{tab:overall_evaluation_results} compares the NMSE, model complexity, and inference efficiency of all methods. All baseline methods produce multi-step predictions using recursive rollout, in which each predicted CIR is reused as the input for the next step. As a result, errors may accumulate across the prediction steps. LLM4UACP achieves the lowest NMSE values on both measured datasets. Compared with STL-Transformer, LLM4UACP reduces the NMSE by 88.53\% on SO-06 and 85.15\% on SO-07. These consistent reductions across both datasets indicate that the gain is not tied to a single channel realization but reflects stronger adaptation to measured UWA channels under different propagation conditions.

Fig.~\ref{fig:so06_prediction} explains the performance gaps in Table~\ref{tab:overall_evaluation_results}. RLS recovers some high-energy early arrivals but adds scattered artifacts, reflecting limits of linear updates in tracking nonstationary multipath evolution. AttLSTMPreNet and CsiPreNet produce overly smooth vertical patterns, suppressing local temporal variations and weakening delayed paths. STL-Transformer preserves more local dynamics than these baselines, but visible deviations persist in dominant and intermediate-delay regions. LLM4UACP better preserves the sparse delay-time CIR structure, including dominant early arrivals and weaker delayed components, explaining the substantial NMSE reduction.

To assess the downstream communication performance of channel prediction, we conduct an OFDM-based UWA transmission simulation. Specifically, we assume that the CSI at previous instants is perfectly acquired, whereas the channel experienced by each OFDM symbol is obtained from prior channel prediction. The multiple consecutive future CIR snapshots jointly predicted by LLM4UACP are applied to the corresponding OFDM symbols. The predicted CIRs are interpolated onto the original delay grid and are then transformed into frequency-domain channel responses via DFT before receiver detection. The OFDM system follows the same parameters as the carrier frequency and sampling frequency in the sounding method with $B\approx5$ kHz and $N_{\rm FFT} = 4096$. The cyclic prefix (CP) is assumed to be sufficiently long to eliminate inter-symbol interference, such that the BER comparison primarily reflects the impact of channel prediction accuracy. A rate-$1/3$ irregular QC-LDPC code over $\mathrm{GF}(2)$ with codeword length 336 and BPSK modulation is adopted. We perform factor-graph detection and LDPC decoding with one and three iterations, respectively.

\begin{figure}[!]
    \centering
    {\includegraphics[width=0.49\linewidth]{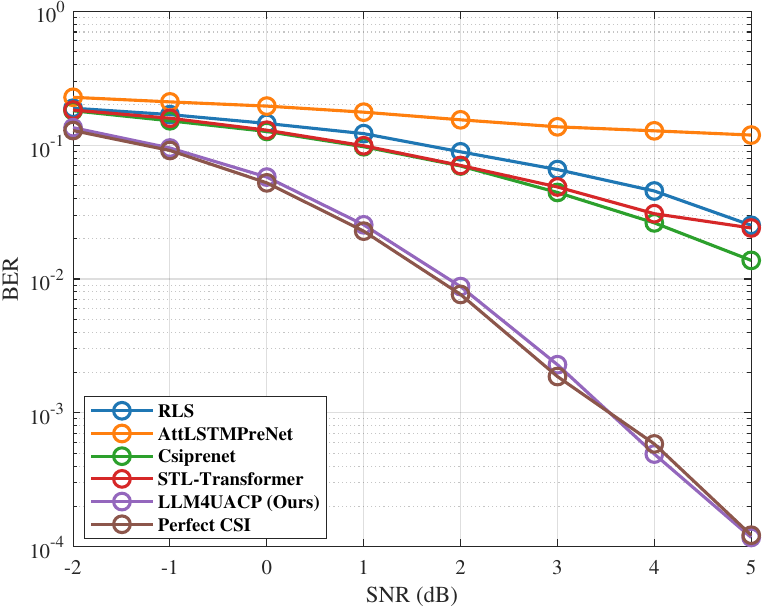}\label{fig:SO06}}
    \hfill
    {\includegraphics[width=0.49\linewidth]{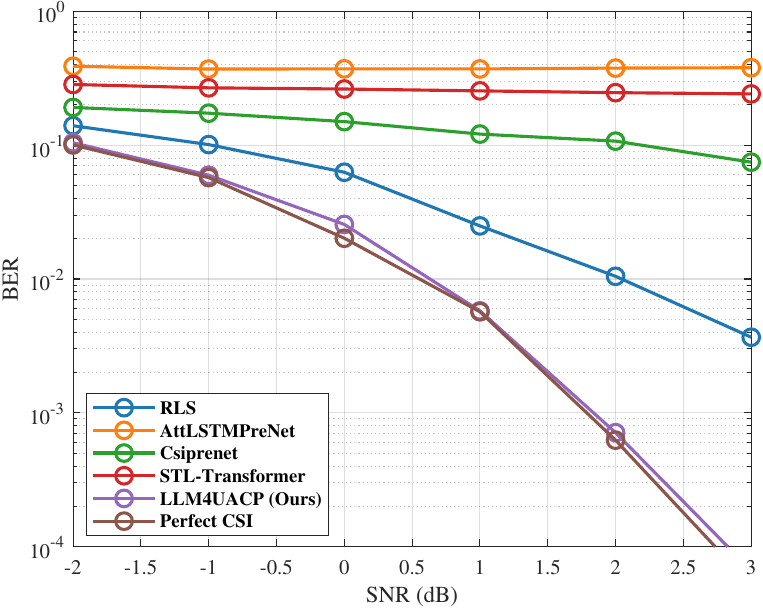} \label{fig:SO07}}
    \caption{BER comparison in SO-06 (left) and SO-07 (right). The predicted channels are used for signal detection, while the true channel serves as the Perfect CSI benchmark.}
    \label{fig:BER}
    
\end{figure}

Fig.~\ref{fig:BER} shows the BER performance when the predicted channels are used for signal detection. On both SO-06 and SO-07, LLM4UACP nearly overlaps with the Perfect CSI benchmark and provides about $1.5$ dB or more SNR gain over the compared methods at a BER of $10^{-2}$, while most baselines fail to reach this BER within the tested SNR range. It is also observed that, for some baselines, the NMSE ranking is not consistent with the BER performance. This mismatch arises because NMSE measures the average channel reconstruction error, whereas BER is more sensitive to communication-relevant structures, such as dominant paths, phase information, and detection-effective channel components. These results demonstrate that LLM4UACP better preserves such structures, leading to superior end-to-end performance.

LLM4UACP incurs most of its computational cost during inference. Although the model has 21.08M parameters, BOFT freezes the backbone and updates only 0.11M adaptation parameters, so its trainable parameter count is comparable to those of lightweight baselines. At inference time, however, the frozen backbone must still be evaluated in full, leading to a higher computational cost than that of the lightweight baselines. The results for UWA channel prediction show that LLM4UACP achieves higher prediction accuracy and better BER performance at this additional inference cost.

\begin{figure}[!t]
\centering
\includegraphics[width=1\columnwidth]{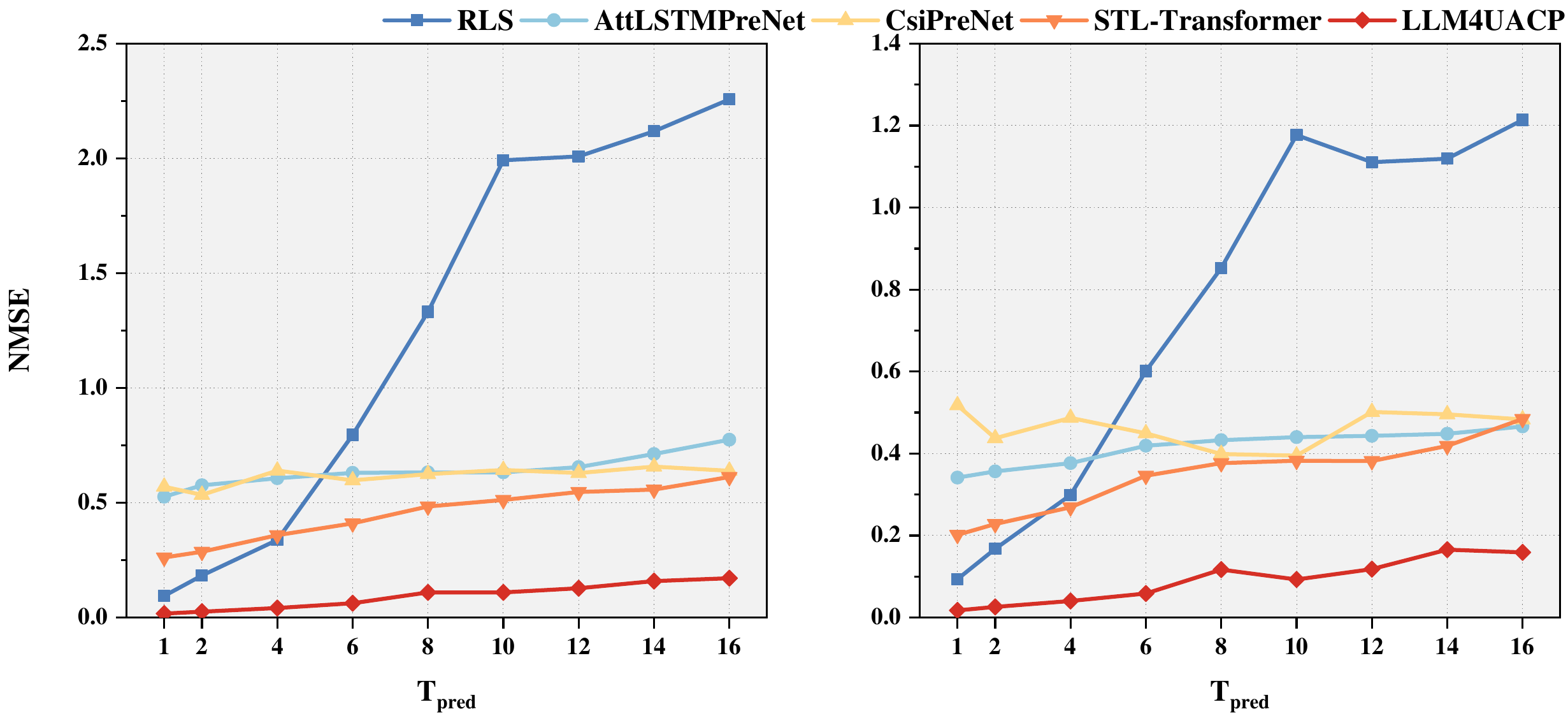}
\caption{NMSE under prediction horizon $T_{pred}$ with fixed $T_{obs}=16$ on SO-06 (left) and SO-07 (right).}
\label{fig:pred_horizon}
\end{figure}

\begin{figure}[!t]
\centering
\includegraphics[width=1\columnwidth]{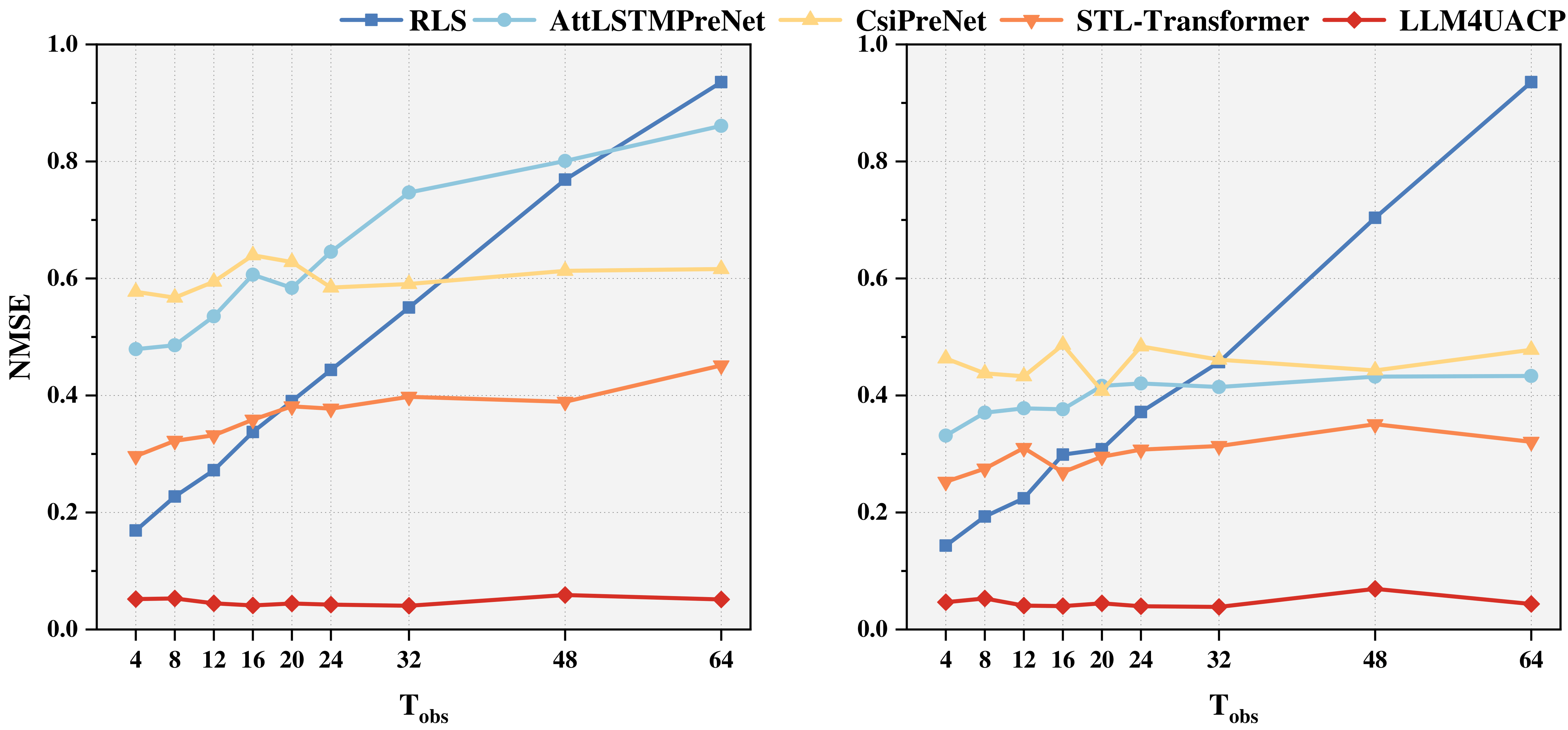}
\caption{NMSE under observation window length $T_{obs}$ with fixed $T_{pred}=4$ on SO-06 (left) and SO-07 (right).}
\label{fig:obs_window}
\end{figure}

\subsection{Impact of Prediction Steps and Observation Window}
\label{subsec:temporal_params}

In practical UWA communications, channel predictor effectiveness largely depends on the prediction steps $T_{pred}$ and observation window $T_{obs}$. Under severe multipath propagation, Doppler distortion, and strong nonstationarity in UWA channels, these parameters affect both prediction accuracy and computational efficiency. However, appropriate settings for UWA channel prediction remain insufficiently studied. We therefore conduct a controlled study, varying one parameter at a time while keeping the other fixed.

As shown in Fig.~\ref{fig:pred_horizon}, the NMSE of most methods increases with $T_{pred}$ because longer horizons weaken the correlation between the observed history and future CIRs. RLS degrades sharply under nonstationary multipath evolution, and the learning-based baselines still fail to fully preserve delay-time structure. LLM4UACP maintains the lowest NMSE and the smallest error growth, indicating stronger robustness to multi-step prediction.

As shown in Fig.~\ref{fig:obs_window}, increasing $T_{obs}$ does not consistently improve prediction accuracy. A short observation window provides limited channel history and may miss the recent evolution of dominant paths, whereas an overly long window may include outdated observations with statistics mismatched to the current UWA channel state. The baseline methods are sensitive to this tradeoff. In contrast, LLM4UACP maintains consistently low prediction errors across different observation windows, suggesting stronger robustness to the balance between insufficient history and outdated channel information.

Overall, the configuration with $T_{obs}=16$ and $T_{pred}=4$ provides a balanced operating point, offering sufficient temporal context for modeling channel evolution while limiting the influence of nonstationary historical observations and avoiding unnecessary computational overhead.

\begin{figure}[!t]
\centering
\includegraphics[width=1\columnwidth]{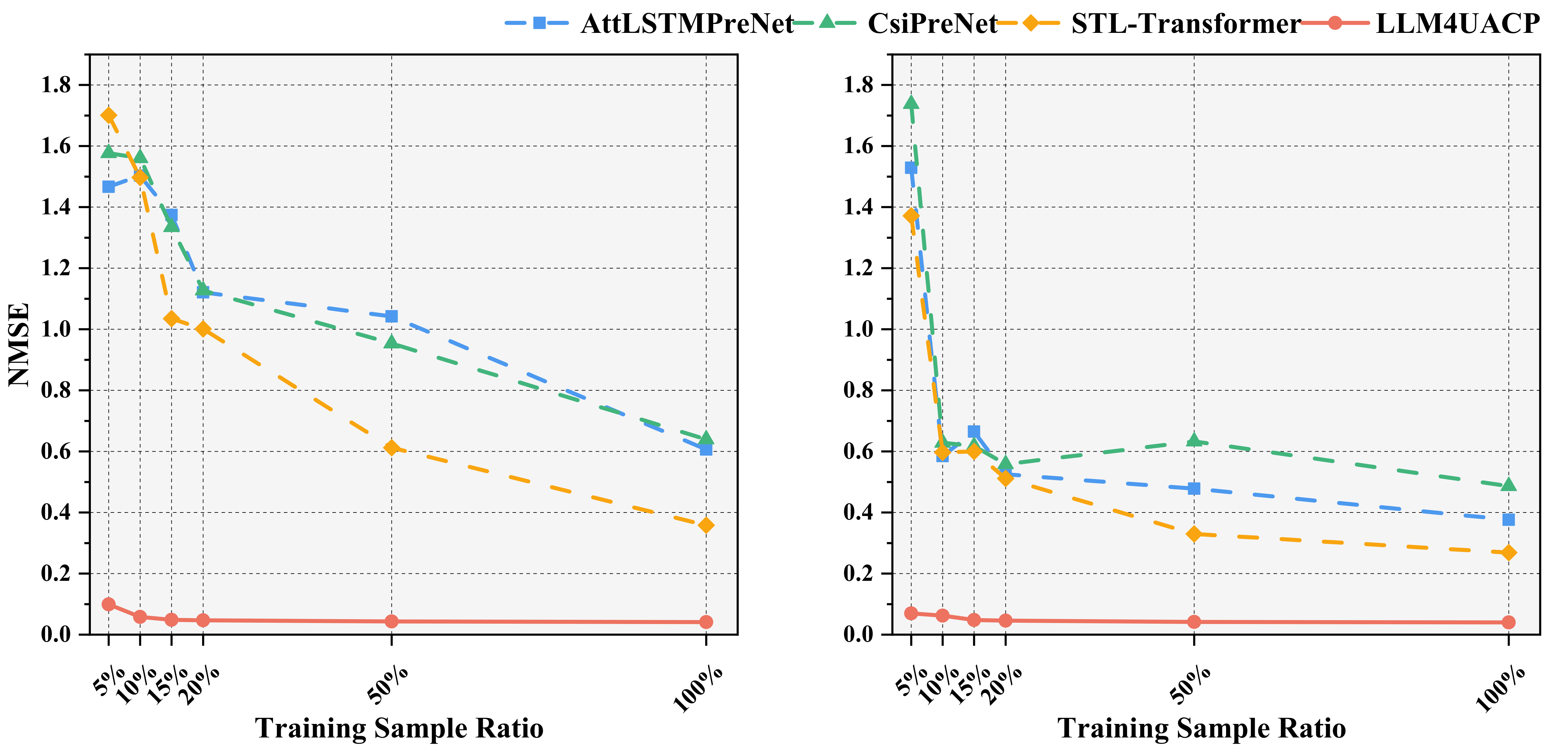}
\caption{Few-shot NMSE performance under different training sample ratios on SO-06 (left) and SO-07 (right).}
\label{fig:few_shot}
\end{figure}

\begin{figure}[!t]
\centering
\includegraphics[width=1\columnwidth]{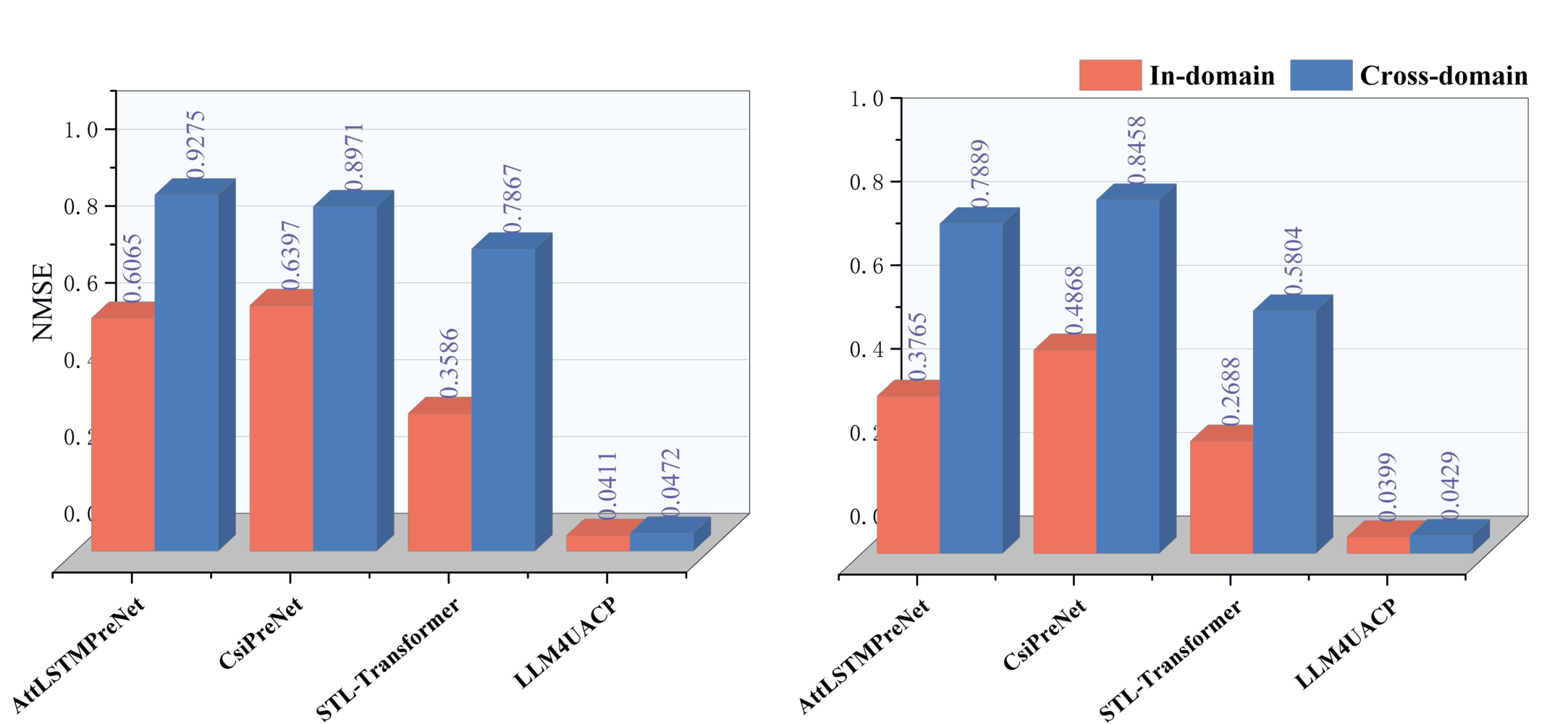}
\caption{Cross-dataset generalization. In-domain denotes same-dataset training and testing, and cross-domain denotes direct target-dataset evaluation without target-domain training. Left: SO-06 to SO-07. Right: SO-07 to SO-06.}
\label{fig:zero_shot}
\end{figure}

\subsection{Data Efficiency and Cross-Dataset Generalization}
\label{subsec:generalization}

Beyond prediction accuracy, practical UWA channel prediction also requires data efficiency and robustness across sea trials. Measured UWA channel data are costly to collect, and channel statistics can vary substantially across different propagation environments. We therefore evaluate LLM4UACP under two practical settings, limited training data and cross-dataset transfer.

Fig.~\ref{fig:few_shot} evaluates data efficiency by varying the training-sample ratio on SO-06 and SO-07. The baseline methods degrade rapidly as the amount of training data decreases, indicating strong dependence on target-domain samples and sensitivity to dataset-specific fluctuations. In contrast, LLM4UACP maintains substantially lower NMSE across all ratios, showing that the pretrained LLM provides useful prior knowledge for temporal channel modeling, while BOFT adapts the model with few trainable parameters and reduces dependence on large measured datasets.

Fig.~\ref{fig:zero_shot} evaluates cross-dataset transfer by training on one sea trial and testing directly on the other. The baseline methods exhibit a large gap between in-domain and cross-domain performance, indicating sensitivity to source-domain propagation statistics. In contrast, LLM4UACP shows only mild degradation, with the NMSE increasing from 0.0411 to 0.0472 when transferring from SO-06 to SO-07 and from 0.0399 to 0.0429 in the reverse direction. This small gap suggests that LLM4UACP captures channel evolution patterns that are less dependent on a specific sea trial and transfer better between the two measured UWA records.
\begin{table*}[!t]
\caption{Grouped Ablation Study of LLM4UACP under High-Resolution Prediction Settings.
$\Delta$ denotes the relative NMSE increase compared with the full model.}
\label{tab:wo_ablation_study}
\centering
\scriptsize
\setlength{\tabcolsep}{4.0pt}
\renewcommand{\arraystretch}{1.42}

\resizebox{\textwidth}{!}{%
\begin{tabular}{
|p{0.15\textwidth}
|p{0.17\textwidth}
|cc
|>{\centering\arraybackslash}p{0.078\textwidth}
 >{\centering\arraybackslash}p{0.078\textwidth}
|>{\centering\arraybackslash}p{0.078\textwidth}
 >{\centering\arraybackslash}p{0.078\textwidth}|
}
\hline
\multirow{2}{0.15\textwidth}{
    \centering
    \rule{0pt}{7.2ex}
    \textbf{\shortstack{Ablation\\Group}}
}
&
\multirow{2}{0.17\textwidth}{
    \centering
    \rule{0pt}{7.2ex}
    \textbf{\shortstack{w/o Ablation\\Variant}}
}
& \multicolumn{2}{c|}{
    \rule[-0.9ex]{0pt}{4.5ex}
    \textbf{Model Scale}
}
& \multicolumn{4}{c|}{
    \rule[-0.9ex]{0pt}{4.5ex}
    \textbf{Prediction Error}
} \\[0.6ex]
\cline{3-8}

&
& \rule[-1.8ex]{0pt}{6.4ex}\textbf{\shortstack{Total\\Params (M)}}
& \rule[-1.8ex]{0pt}{6.4ex}\textbf{\shortstack{Trainable\\Params (M)}}
& \rule[-1.8ex]{0pt}{6.4ex}\textbf{\shortstack{NMSE\\(SO-06)}}
& \rule[-1.8ex]{0pt}{6.4ex}\textbf{\shortstack{$\Delta$\\(SO-06)}}
& \rule[-1.8ex]{0pt}{6.4ex}\textbf{\shortstack{NMSE\\(SO-07)}}
& \rule[-1.8ex]{0pt}{6.4ex}\textbf{\shortstack{$\Delta$\\(SO-07)}} \\
\hline

\multirow{1}{0.15\textwidth}{
    \centering
    \textbf{\shortstack{Full Model}}
}
& \centering\textbf{\mbox{LLM4UACP}} 
& 21.08 & 0.11 & 0.0411 & -- & 0.0399 & -- \\
\hline

\multirow{5}{0.15\textwidth}{
    \centering
    \textbf{\shortstack{Dual-Domain\\Feature\\Extraction}}
}
& \centering\mbox{w/o Frequency Branch} 
& 21.07 & 0.10 & 0.0442 & +7.6\% & 0.0412 & +3.3\% \\
\cline{2-8}

& \centering\mbox{w/o Delay Branch} 
& 21.07 & 0.10 & 0.0505 & +22.3\% & 0.0474 & +18.9\% \\
\cline{2-8}

& \centering\mbox{w/o Linear Bypass} 
& 21.08 & 0.11 & 0.0581 & +41.4\% & 0.0577 & +44.7\% \\
\cline{2-8}

& \centering\mbox{w/o Adaptive Fusion} 
& 21.08 & 0.11 & 0.0427 & +3.9\% & 0.0413 & +3.6\% \\
\cline{2-8}

& \centering\mbox{w/o IDFT} 
& 21.08 & 0.11 & 0.0430 & +4.7\% & 0.0415 & +4.1\% \\
\hline

\multirow{1}{0.15\textwidth}{
    \centering
    \textbf{\shortstack{Tokenization}}
}
& \centering\mbox{w/o Delay-Time Patch} 
& 21.32 & 0.34 & 0.1874 & +355.9\% & 0.1388 & +247.9\% \\
\hline

\multirow{3}{0.15\textwidth}{
    \centering
    \textbf{\shortstack{Pretrained Backbone}}
}
& \centering\mbox{w/o pretraining} 
& 21.00 & 21.00 & 0.0509 & +23.9\% & 0.0437 & +9.6\% \\
\cline{2-8}

& \centering\mbox{w/ Llama-3.2-1B} 
& 628.00 & 0.40 & 0.0428 & +4.1\% & 0.0422 & +5.8\% \\
\cline{2-8}

& \centering\mbox{w/ Qwen2.5-0.5B} 
& 225.78 & 0.17 & 0.0443 & +7.8\% & 0.0420 & +5.3\% \\
\cline{2-8}

\hline

\multirow{3}{0.15\textwidth}{
    \centering
    \textbf{\shortstack{PEFT}}
}
& \centering\mbox{w/ full fine-tuning} 
& 21.00 & 21.00 & 0.0496 & +20.7\% & 0.0465 & +16.6\% \\
\cline{2-8}

& \centering\mbox{w/ $\text{LoRA}_{r=16}$\cite{LoRA}} 
& 21.30 & 0.32 & 0.0422 & +2.6\% & 0.0411 & +3.1\% \\
\cline{2-8}

& \centering\mbox{w/ FPT\cite{liu2024llm4cp}} 
& 21.00 & 2.13 & 0.0419 & +2.0\% & 0.0394 & -1.3\% \\
\hline

\multirow{2}{0.15\textwidth}{
    \centering
    \textbf{\shortstack{Loss}}
}
& \centering\mbox{w/o gradient loss} 
& 21.08 & 0.11 & 0.0426 & +3.7\% & 0.0409 & +2.6\% \\
\cline{2-8}

& \centering\mbox{w/o adaptive weight} 
& 21.08 & 0.11 & 0.0422 & +2.7\% & 0.0422 & +5.8\% \\
\hline

\end{tabular}%
}
\end{table*}

\subsection{Ablation Studies}
\label{subsec:architecture_analysis}

Table~\ref{tab:wo_ablation_study} evaluates the main architectural designs in LLM4UACP. The full model uses only 0.11M trainable parameters and achieves the best overall accuracy-parameter tradeoff, indicating that the gain comes from how UWA channel structure is represented and adapted rather than from trainable parameter count alone.

The ablation study on tokenization shows that delay-time structured tokens are essential for adapting the pretrained backbone to UWA channel modeling. Removing the proposed patch embedding leads to the largest performance degradation, increasing the NMSE by $355.9\%$ on SO-06 and $247.9\%$ on SO-07. This result confirms that grouping neighboring delay taps and consecutive snapshots before they are fed into the backbone is critical for capturing correlated multipath evolution.

The backbone comparison indicates that pretraining provides useful sequence priors, whereas larger backbones do not necessarily improve CIR prediction. Removing pretrained initialization increases the NMSE by 23.9\% on SO-06 and 9.6\% on SO-07, while replacing the adopted backbone with other LLMs yields no consistent improvement. These results suggest that effective CIR prediction depends on presenting channel evolution to the backbone in a structured form aligned with its sequence modeling capability.

The dual-domain ablation results show that a single-domain input omits useful channel information. Removing the delay branch causes larger NMSE increases than removing the frequency branch, while removing IDFT alignment or adaptive fusion also degrades performance. These results indicate that the CIR carries the main predictive structure, the CFR provides complementary frequency-selective cues, and these cues should be delay-aligned and adaptively fused. The linear bypass also matters. Its removal increases the NMSE by 41.4\% on SO-06 and 44.7\% on SO-07, showing that the short-term linear channel trend remains useful for prediction.

The ablation results for adaptation and loss design show that these components mainly enhance robustness. Full fine-tuning yields worse performance despite updating all backbone parameters, suggesting over-adaptation when measured data are limited. Compared with LoRA and FPT, BOFT provides a better accuracy-parameter tradeoff by preserving the pretrained representation while adjusting token interactions in the attention space. The smaller performance gaps in the loss ablation indicate that temporal-gradient consistency and adaptive weighting mainly stabilize multi-step prediction.

\section{Conclusion And Future work}
\label{sec:conclusion}

This paper studied multi-step CIR prediction for UWA communication under CSI aging, time-varying multipath propagation, and limited sea-trial data. We proposed LLM4UACP, a structure-preserving pretrained Transformer framework that predicts multiple future CIR snapshots in one forward pass. It connects measured UWA channels to a frozen pretrained sequence model via dual-domain CIR/CFR feature extraction, Fourier-consistent delay-domain alignment, delay-time grid embedding, and BOFT-based parameter-efficient adaptation.

Experiments on SO-06 and SO-07 sea-trial datasets show that LLM4UACP enables accurate, data-efficient multi-step prediction. Under $(T_{\mathrm{obs}},T_{\mathrm{pred}})=(16,4)$, LLM4UACP achieves NMSE values of 0.0411 and 0.0399, corresponding to reductions of 88.53\% and 85.15\% over the latest baseline method, respectively. In the simulated OFDM link, the predicted channels also yield BER close to the Perfect CSI benchmark in the tested OFDM simulator. Robustness and ablation studies further confirm the roles of delay-time structure preservation, dual-domain physical alignment, the linear bypass, and BOFT-based adaptation.

Future work will improve the deployment feasibility of LLM4UACP in resource-constrained UWA systems. Although the current model achieves high prediction accuracy, its large parameter count and inference cost limit deployment on many edge devices. Future efforts will therefore investigate lightweight model compression methods, such as quantization, knowledge distillation, and structured pruning, to reduce computational and memory demands while maintaining prediction performance.

\bibliographystyle{IEEEtran} 
\bibliography{ref}   

\vfill

\end{document}